\documentclass{statsoc}

\usepackage{graphicx}
\usepackage{natbib,amsfonts,amsmath,verbatim}
\usepackage{amssymb}
\usepackage{rotating}
\usepackage{graphics,psfrag} 

\newcommand{\append}{Appendix~}
\newcommand{\sect}{Section~}

\newcommand{\chaps}{Chapters~}
\newcommand{\fig}{Figure~}

\newcommand{\MCMC}{\ensuremath{\text{Markov chain Monte Carlo}}}

\newcommand{\Sigmahat }{\ensuremath{\widehat \Sigma }}

\newcommand{\ie}{\ensuremath{\text{i.e. }}}
\newcommand{\half}{\ensuremath{\frac{1}{2}}}
\newcommand{\etr}{\ensuremath{\text{etr}}}
\newcommand{\trace}{\ensuremath{\text{trace}}}

\newcommand{\eg}{\ensuremath{\text{e.g. }}}

\newcommand{\MH}{\ensuremath{{\text{Metropolis Hastings}}}}
\newcommand{\offdiagonal}{\ensuremath{{\text{off-diagonal}}}}
\newcommand{\nonzero}{\ensuremath{{\text{nonzero}}}}

\newcommand{\nonGaussian}{\ensuremath{{\text{non-Gaussian}}}}
\newcommand{\Vbar}{\ensuremath{\overline V }} 
 
\newcommand{\Vwbaratio}{\ensuremath{\overline V_w (l)  /\overline V_w(l+1) }} 
\newcommand{\Vdbaratio}{\ensuremath{\overline V_d (l)  /\overline V_d(l+1) }} 
\newcommand{\nc}{\ensuremath{\text{normalizing constants}}}

\newcommand{\mle}{\ensuremath{\text{maximum likelihood estimate}}}

\newcommand{\ra}{\ensuremath{\rightarrow}}

\newcommand{\bmm}{\ensuremath{\text{bmm}}}

\newcommand{\Jhat }{\ensuremath{\widehat J }}
\newcommand{\Omegahat }{\ensuremath{\widehat \Omega }}

\newtheorem{theorem}{Theorem}

\newtheorem{lemma}[theorem]{Lemma}

\title[Bayesian Covariance Matrix Estimation]{Bayesian Covariance Matrix Estimation using a Mixture of Decomposable Graphical Models}
\author[Armstrong {\it et al.}]{Helen Armstrong}
\address{School of Mathematics, University of New South Wales
	Sydney,
	Australia.}
\email{helen@maths.unsw.edu.au}
\author{Christopher K. Carter}
\address{Australian School of Business, University of New South Wales,
	Sydney,
	Australia.}
\author{Kevin K. F. Wong}
\address{The Institute of Statistical Mathematics,
	Graduate University for Advanced Studies,
	Tokyo,
 	Japan.}
\author[Armstrong {\it et al.}]{Robert Kohn}
\address{Australian School of Business, University of New South Wales,
	Sydney,
	Australia.}
\begin{document}

\maketitle
\begin{abstract}
Estimating a covariance matrix efficiently and discovering its structure
are important statistical problems with applications in many fields.
This article takes a Bayesian approach to estimate the covariance matrix
of Gaussian data.
We use ideas from Gaussian graphical models and model selection to construct
a prior for the covariance matrix that is a mixture over all
decomposable graphs,
where a graph means the configuration of nonzero off-diagonal elements
in the inverse of the covariance matrix.
Our prior for the covariance matrix is such that the probability of each
graph size is specified by the user and graphs of equal size are assigned
equal probability.
Most previous approaches assume that all graphs are equally probable.
We give empirical results that show the prior that assigns equal probability
over graph sizes outperforms the prior that assigns equal probability over
all graphs, both in identifying the correct decomposable graph and in
more efficiently estimating the covariance matrix.
The advantage is greatest when the number of observations is small relative
to the dimension of the covariance matrix.
Our method requires the number of decomposable graphs for each graph size. 
We show how to estimate these numbers using simulation and that the
simulation results agree with analytic results when such results are known.
We also show how to estimate the posterior distribution of the covariance
matrix using Markov chain Monte Carlo with the elements of
the covariance matrix integrated out and give empirical results that
show the sampler is much more efficient than current methods.
The article also shows empirically that there is minimal change in
statistical efficiency in using the mixture over decomposable graphs 
prior for estimating
a general covariance compared to the Bayesian estimator by \cite{Wong_C_K03},  
even when the graph of the covariance matrix is nondecomposable.
However, our approach has some important computational advantages over that of
\cite{Wong_C_K03}.

Finally, we note that both the prior and the simulation method to evaluate
the prior apply generally to any decomposable graphical model. 


\end{abstract}

\section{Introduction\label{S:intro}} 
Estimating a covariance matrix efficiently is an important statistical problem
with many applications, such as multivariate regression, cluster analysis,
factor analysis, and discriminant analysis; 
see, for example, \cite{Mardia_K_B79}.
Such applications are used in the fields of Business, Engineering,
and the physical and social sciences. 
It is also of considerable interest to understand the graphical structure
of the covariance matrix because it is directly interpretable in terms
of the partial correlations of the underlying multivariate distribution.
By the graph of the covariance matrix we mean the pattern of nonzero off diagonal elements
in the inverse of the covariance matrix, also called the concentration matrix
(see \citealt[Chapter~5]{Lauritzen96}).
Estimating a covariance matrix efficiently and understanding its graphical structure
are difficult estimation problems because the number of unknown parameters in the
covariance matrix increases quadratically with dimension and by the requirement that
the estimate of the covariance matrix is positive definite. 

There is a large literature of methods that use shrinkage or Bayesian models 
to improve on the maximum likelihood estimator of the covariance matrix. 
See, for example, \cite{Dempster69}, \cite{Dempster72}, \cite{Efron_M76},
\cite{yang94}, \cite{chiu96}, \cite{Giudici_G99},
\cite{barnard00}, \cite{Wong_C_K03} and \cite{Liechty_L_M04}.
The simulation studies in \cite{yang94} and \cite{Wong_C_K03} show that
considerable gains in efficiency are possible.

\cite{Dempster72} advocates a covariance selection approach to estimate
a covariance matrix more efficiently, by which he means setting to zero
some of the off-diagonal elements of the concentration matrix.
His idea is that a more parsimonious model will give greater efficiency.
However, the selection of which elements to set to 
zero is difficult even for moderate dimensions because a $p\times p$
concentration matrix has $p(p-1)/2$ distinct off-diagonal entries and
there are $2^{p(p-1)/2}$ possible graphs associated with it. 
\cite{Drton_P04} give a model selection approach based on simultaneous 
confidence intervals to determine which partial correlations are zero. 
The simultaneous
confidence intervals are based on large sample theory and become large
when $p$ is moderate to large. \cite{Drton_P04} do not attempt to
estimate the covariance matrix based on their selected graph.

A number of articles take a Bayesian approach to covariance selection.
For the case of decomposable graphs,
\cite{Dawid_L93} introduces a conjugate prior for the covariance
matrix called the hyper inverse Wishart distribution.
\cite{Giudici96} uses a prior for the covariance matrix that is
a mixture of fixed parameter hyper inverse Wishart priors
over decomposable graphs and calculates the marginal likelihood for each
decomposable graph. The marginal likelihood is used to calculate the posterior probability of each graph. 
This gives an exact solution for small examples, but 
for $p$ greater than approximately 8 the number of graphs is prohibitively large. 

\cite{Roverato00} shows that the hyper inverse Wishart prior for the covariance matrix
is equivalent to a constrained Wishart prior for the concentration matrix.
Although it is straightforward to define a constrained Wishart prior for general graphs,
such distributions have \nc{} that are not available analytically
unless the graph is decomposable.
\cite{Roverato02}, \cite{Dellaportas_G_R04} and \cite{Kayis_M} 
propose efficient simulation and importance sampling methods for estimating
the normalizing constants for the nondecomposable graphs.
The normalizing constants are used to examine a small number of graphs
and select those that that have the
highest marginal likelihood or posterior probability,
rather than to estimate the covariance matrix by averaging over graphs.
However, such an approach seems unsuitable as the basis of a \MCMC{} sampling scheme
when $p$ is moderate to large because there are $2^{p(p-1)/2}$ possible graphs
with only a small fraction of them being decomposable. 

\cite{Giudici_G99} give a MCMC approach that can deal with large values of $p$. 
Their method applies to a hierarchical model with a hyper inverse Wishart prior for the 
covariance matrix conditional on a decomposable graph. 
They use reversible jump Metropolis-Hastings methods to generate the covariance matrix 
and other parameters.
Their method has a local computation property that only requires 
Cholesky decompositions of the submatrix of the covariance matix corresponding 
to a clique of the graph. 
\cite{Brooks_G_R03} modify the reversible jump MCMC proposal of \cite{Giudici_G99}
and give empirical results to show this improves the convergence rate.

\cite{Wong_C_K03} also use MCMC methods to select which off-diagonal element
to set to zero. 
They use reversible jump Metropolis-Hastings methods to generate the 
inverse covariance matrix and other parameters. 
The main difference between \cite{Giudici_G99} and \cite{Wong_C_K03}
is that \cite{Wong_C_K03} do not constrain the possible graphs to be decomposable. 
\cite{Wong_C_K03} use a prior with normalizing constants based on graph size
to avoid having to calculate normalizing constants for each nondecomposable graph.
They also need to run a separate MCMC to estimate the normalizing constants
for each graph size.

For longitudinal data, \cite{Smith_K02} factor the concentration matrix
using a Cholesky decomposition and carry out variable selection on the strict
lower triangle of the Cholesky to obtain parsimony.
Their approach is attractive when there is some natural ordering of the observation vector,
but there are two potential drawbacks to the Cholesky approach when such
a natural ordering does not exist.
First, different orderings of the variables can yield different estimates
of the covariance matrix.
Second, under some orderings the Cholesky factor may be quite full even
if the concentration matrix is sparse. 

In this paper we consider Bayesian estimation of decomposable covariance selection models,
also known as decomposable graphical Gaussian models. 
Our article makes the following contributions. 
First, we propose a prior for the covariance matrix such that the 
probability of each graph size is specified by the user,
whereas most previous approaches, e.g. \cite{Giudici_G99}, 
assume that all graphs are equally probable.
We show by simulation that the prior that assigns 
equal probability over graph sizes outperforms the prior that assigns equal probability
over all graphs, both in identifying the correct decomposable model and in estimating the covariance matrix more efficiently. 
This advantage is greatest when the number of observations is small relative
to the dimension of the covariance matrix.
We also show by simulation that there is minimal change in statistical efficiency
in using our mixture prior compared to the estimator of \cite{Wong_C_K03},
even when the graph of the covariance matrix is nondecomposable. 

Our prior requires knowing the number of decomposable graphs for each graph size.
The second contribution of the article is to give a MCMC method for estimating these counts. 
and to show that the counts obtained by the simulation method agree 
with analytic results when such results are known. 

Our third contribution is to use the marginal likelihood results in \cite{Giudici96} to 
derive a reduced conditional MCMC sampler for decomposable graphical models, 
where the covariance matrix is integrated out of all conditional distributions 
and is not generated in the MCMC.
Our approach does not require reversible jump Metropolis-Hasting methods
and has the local computation properties of the \cite{Giudici_G99} approach, 
so the computational complexity for one iteration of our approach is similar to that of 
\cite{Giudici_G99}. 
We give empirical results that show our sampler produces iterates that have much less
autocorrelation compared to the methods in \cite{Brooks_G_R03}.
We also show that our sampler has a faster convergence rate than the \cite{Wong_C_K03}
approach. 
\cite{Jones05} uses a version of the marginal likelihood MCMC approach described
in this paper that does not involve hyperparameters.
This approach is used to find the graph with maximum posterior probability
and the results are compared to stochastic search. 

The results in our article suggest that at present there is no \lq best\rq{} 
 method for estimating Gaussian covariance selection models.
While the method of \cite{Wong_C_K03} works in principle for all graphs,
the convergence of their MCMC simulation can be slow if the true graph
has full subgraphs of size 5 or larger because \cite{Wong_C_K03} generate the elements of
the concentration matrix one at a time.
On the other hand the sampling scheme for decomposable graphs presented in our article
is extremely efficient because the concentration matrix is integrated out and is an
attractive alternative to the \cite{Wong_C_K03} model for high dimensional graphs
that are likely to have substantial full subgraphs. 
There are two other advantages of the decomposable prior considered in our article.
The first is that there is a separate normalizing constant for each decomposable graph,
whereas \cite{Wong_C_K03} have a normalizing constant for each graph size.
The second is that the \cite{Wong_C_K03} model does not at present allow for
hyperparameters in the prior. For example, using an equicorrelated prior as in 
\cite{Giudici_G99} is not at present feasible with the approach of \cite{Wong_C_K03}.

The paper is organized as follows.
Section \ref{ggm} briefly introduces graphical Gaussian models. 
Section \ref{sec:model} describes our Baysian covariance selection model and 
Section \ref{sec:sampling} describes our MCMC approach to estimating this model. 
Section \ref{sec:size_vs_unif} compares the prior that assigns equal probablity to
each graph size to the prior that assigns equal probability to each decomposable graph.
Section \ref{sec:counting} shows how to estimate the number of decomposable graphs for
each size by simulation.
Section \ref{sec:us_vs_BGR} compares the efficiency of our sampler to
the reversible jump approach in \cite{Brooks_G_R03}.
Section \ref{sec:physical_p11responses_no_covariates} gives a Bayesian analysis
of a multivariate dataset on physical measurements.
Section \ref{sec:us_vs_ed} compares the prior that assigns equal probability to
each graph size to the prior of \cite{Wong_C_K03}.
There are two appendices.
The first gives the proofs of the results in the paper.
The second gives a computationally efficient expression for evaluating
the ratio of normalizing constants from Section~\ref{gen:g}.

\section{Background on Gaussian graphical models\label{ggm}}
Before explaining our Bayesian covariance selection model we
provide some background on Gaussian graphical models.
Further details on such models are available in \cite{Dawid_L93} and 
\chaps 2, 3 and 5 of \cite{Lauritzen96}. 

Let $g=(V,E)$ be an undirected graph with vertices $V=\{1,\ldots,p\}$ and set of
edges $E \subseteq V \times V $. 
For a square matrix $A$ we write $A>0$ to denote that $A$ is positive
definite. Let $M^{+}(g)$ be the set of $p\times p$ matrices $\Omega$ satisfying
$\Omega>0$ and $\Omega_{ij}=0$ for all pairs $(i,j) \notin E$. 

For a 
given $p \times p$
covariance matrix $\Sigma$, we 
define the \textit{graph} of $\Sigma$, $g=g(\Sigma)=(V,E)$, 
as follows. Let 
$\Omega=\Sigma^{-1}$. Let $V=\{1,\ldots,p\}$ and define 
$E=\{(i,j), i \neq j \text{ such that } \Omega_{ij} \neq 0 \}$.
Thus the graph $g=g(\Sigma)$ gives the configuration of nonzero
off-diagonal elements in $\Omega$. 
 
We say that an $m \times m$ matrix $A>0$ has an inverse
Wishart (IW) density with $\delta >0$ degrees of freedom and scale matrix
$\Phi$, denoted as $A \sim IW(m,\delta, \Phi),$ if the density of $A$ is
\begin{equation}
p(A | \delta, \Phi) = \frac{|\frac{\Phi}{2}|^{\frac{\delta}{2}}}{\Gamma_m(\frac{\delta}{2})} 
|A|^{-\frac{(\delta + m +1)}{2}}
\operatorname*{etr} \left(  -\frac{1}{2} \Phi A^{-1} \right), \label{eqn:iw}
\end{equation}
where $\operatorname*{etr}(A)=\exp(\operatorname{trace}(A))$
and for $\alpha > \frac{(m-1)}{2},$ 
\[
\Gamma_m(\alpha)= \pi^{m(m-1)/4} \prod_{i=1}^m \Gamma(\alpha - (\frac{i-1}{2}) )
\]
is the multivariate gamma function \citep[p.~113]{Muirhead82}.

Lauritzen (1996, Definition~2.3, p.8) defines a 
\textit{decomposable} graph and we refer to a covariance matrix $\Sigma$ 
as decomposable if its graph $g = g(\Sigma)$ is decomposable. 

Suppose that $g$
is a decomposable graph and let $C_{1},\ldots,C_{k}$ be a perfect sequence of
the cliques of $g$. Let $H_{j}=C_{1}\cup\ldots\cup C_{j}$ be the history of
the sequence and let $S_{j}=H_{j-1}\cap C_{j}$ be the separators for
$j=2,\ldots,k$.
For any matrix
$M$ and subset of vertices $B$, 
use $M_{BB}$ to denote the symmetric submatrix of $M$ which is 
formed by taking every corresponding entry $M_{ij}$ for 
which the vertices $\{V_i,V_j \} \in B$.
Using the parameterization of \cite{Dawid81}, we say $\Sigma$ has 
a \textit{hyper inverse Wishart} (HIW) distribution, with 
\textit{hyperparameters} $(\delta,\Phi)$ denoted by
$\Sigma\sim HIW(g,\delta,\Phi)$, if for $\Sigma^{-1}\in M^{+}(g)$%
\begin{equation}
p(\Sigma|\delta,\Phi,g)=\frac{%
{\displaystyle\prod_{i=1}^{k}}
p\left(  \Sigma_{C_{i}C_{i}}|\delta,\Phi_{C_{i}C_{i}}\right)  }{%
{\displaystyle\prod_{i=2}^{k}}
p\left(  \Sigma_{S_{i}S_{i}}|\delta,\Phi_{S_{i}S_{i}}\right)  },
\label{eqn_hiwd_1}%
\end{equation}
where $\delta>0$, $\Phi>0$, and the density is with respect 
Lebesgue measure on the elements of  
$\Sigma$ corresponding to edges of $g$.

In (\ref{eqn_hiwd_1}), the terms 
$p\left(  \Sigma_{C_{i}C_{i}}|\delta,\Phi_{C_{i}C_{i}}\right)$
denote the IW densities 
$\Sigma_{C_{i}C_{i}} \sim IW \left( |C_i|, \delta + |C_i| -1, \Phi_{C_{i}C_{i}} \right)$
given by %
\begin{equation}
p\left(  \Sigma_{C_{i}C_{i}}|\delta,\Phi_{C_{i}C_{i}}\right)  = \frac{\left|
\frac{\Phi_{C_{i}C_{i}}}{2}\right|  ^{\left(  \frac{\delta+\left|
C_{i}\right|-1  }{2}\right)  }}{\Gamma_{\left|  C_{i}\right|  }\left(
\frac{\delta+\left|  C_{i}\right| -1 }{2}\right)  }\left|  \Sigma_{C_{i}C_{i}%
}\right|  ^{-\left(  \frac{\delta+2\left|  C_{i}\right|  }{2}\right)
}\operatorname*{etr}\left[  -\frac{1}{2}\left(  \Sigma_{C_{i}C_{i}}\right)
^{-1}\Phi_{C_{i}C_{i}}\right],  \label{eqn_iwd_2}%
\end{equation}
where $|C_i|$ denotes the cardinality of the clique $C_i$, and the terms $p\left(  \Sigma_{S_{i}S_{i}}|\delta,\Phi_{S_{i}S_{i}}\right)  $
are defined similarly. Note that the expression in (\ref{eqn_hiwd_1}%
) is invariant to the choice of perfect sequence.

From (\ref{eqn:iw}) -- (\ref{eqn_iwd_2}), the normalizing constant for
the HIW distribution is%
\begin{equation}
h(g,\delta,\Phi)=\frac{%
{\displaystyle\prod_{i=1}^{k}}
\left[  \left|  \frac{\Phi_{C_{i}C_{i}}}{2}\right|  ^{ ( \frac
{\delta+\left|  C_{i}\right| -1  }{2} )  }\Gamma_{\left|  C_{i}\right|
}\left(  \frac{\delta+\left|  C_{i}\right| -1  }{2}\right)  ^{-1}\right]  }{%
{\displaystyle\prod_{i=2}^{k}}
\left[  \left|  \frac{\Phi_{S_{i}S_{i}}}{2}\right|  ^{ (  \frac
{\delta+\left|  S_{i}\right| -1 }{2} ) }\Gamma_{\left|  S_{i}\right|
}\left(  \frac{\delta+\left|  S_{i}\right|-1}{2}\right)  ^{-1}\right]  }.
\label{eqn_hiwd_2}%
\end{equation}

\section{Bayesian Covariance Selection models}\label{sec:model}
\subsection{Likelihood and hierarchical structure}\label{subsec:model}
Suppose we have independent observations 
\begin{equation}
y_t \sim  N(\mu, \Sigma),\quad t=1, \cdots, n, 
\end{equation}
where $y_t  $ is $p \times 1.$ Let $y=( y_1, \cdots, y_n)$ be the data.
We use a hierarchical prior for $\mu$ and $\Sigma$ of the form
\[
p( \mu, \Sigma, \Phi, \delta, g )=p(\mu|\Sigma, \Phi, \delta, g)p(\Sigma|\Phi,  \delta, g  )p(\Phi|\delta, g)p(\delta| g)p(g),\
\] where each of the terms on the right is discussed below.
In our article 
we assume that $p(\mu|\Sigma, \Phi, \delta, g) \propto $ constant, as our 
focus is on priors for $\Sigma.$ 
The prior for $\Sigma$ depends on its graph $g$, the $p \times p$ 
matrix $\Phi $ and the scalar $\delta$, and is is discussed
in \sect\ref{prior:sigma}.  
\sect\ref{ggm} defines the graph of $\Sigma$ as 
the configuration of nonzero
off diagonal elements in $\Sigma^{-1}$. 
The prior for $\Phi$ is discussed in \sect\ref{prior:phi} and 
the prior for the graph
$g$ is discussed in \sect\ref{prior:g}.  

In the article we restrict the graph 
of $\Sigma$ to be decomposable, so that
the prior for $\Sigma$ is a mixture over all decomposable 
graphs.
We explain in \sect\ref{ggm} that this is 
equivalent to the prior for $\Omega=\Sigma^{-1}$ 
being a mixture over all Wishart distributions constrained to 
decomposable graphs. 

\subsection{Prior for $\Sigma$}\label{prior:sigma}
We use the HIW prior (\ref{eqn_hiwd_1}) 
for $\Sigma | \Phi, \delta, g$, which allows $\Sigma$ 
to be integrated out in the sampling 
scheme described in \sect\ref{sec:sampling}.
Thus, our prior 
\[
p(d\Sigma|\Phi, \delta)=\sum_g p(d\Sigma|g, \Phi, \delta)p(g)
\]
is a mixture of HIW distributions over all decomposable graphs $g$
As discussed in the introduction, \cite{Roverato00} shows that the inverse of a
HIW random matrix has a Wishart distribution,
subject to the constraints imposed by the corresponding graph. 
Thus 
\[
p(d\Sigma|\Phi, \delta)=\sum_g p(d\Sigma|g, \Phi, \delta)p(g)
\]
is a mixture of of constrained Wishart distributions over all decomposable graphs. 

In our article we set the degrees of freedom parameter $\delta$
to $5$ as such a value of $\delta$ gives a suitably 
noninformative prior for $\Sigma$.

\subsection{Prior specification for \ensuremath{\Phi} and its parameters}\label{prior:phi}
We consider the following three specifications for the hyperparameter 
$\Phi$, and refer to them as the hyperprior forms of $\Phi$:

\begin{enumerate}
\item $\Phi=\tau I, \text{ } \tau >0 $ 
where $I$ is the $p \times p$ identity matrix. 

\item $\Phi=\tau(\rho J + (1-\rho)I), \text{ } \tau >0 $ 
where  $J$ 
is the $p \times p$ matrix of ones and 
$\rho$ is a correlation coefficient that 
needs to be in the open interval
$(-1/(p-1), 1)$ 
for $\Phi$ to be positive definite. This specification is 
used by \cite{Giudici_G99} and is called the \textit{equicorrelated 
version} of $\Phi$ because $\Phi_{ii}=\tau$ and $\Phi_{ij}=\tau \rho$
for $i \neq j$.
 
\item $\Phi =\tau S_y/(n-1), \text{where } \tau >0$, 
\begin{equation} S_y =\sum_{t=1}^n (y_t-\overline{y})(y_t-\overline{y})',\label{eqn:Sy}
\end{equation}
and $\overline{y}$ is the mean of the $y_t$.

We motivate the choice of $\Phi$ in two ways. First, by integrating 
$\mu$ out of $p(y|\mu,\Sigma)$, with $p(\mu)$ constant, we obtain
\begin{equation}
p(y|\Sigma) \propto |\Sigma|^{-(n-1)/2} 
\operatorname*{etr} \left( -\frac{1}{2} S_y \Sigma^{-1} \right).\label{eqn_like_1}
\end{equation} 

Suppose $g$ is a decomposable graph. 
If we take 
$p(\Sigma|g) \propto  p(y|\Sigma)^{1/(n-1)}$,
then from (\ref{eqn_like_1}) and equation (3) of
\cite{Giudici96}, we can write $p(\Sigma|g)$ in the
form (\ref{eqn_iwd_2}) with $\Phi=S_y/(n-1)$.

A second motivation for this choice of $\Phi$ is to note 
that if $\Sigma \sim HIW(p,\delta, \Phi)$, then 
$E(\Sigma_{CC})=\Phi_{CC}/(\delta-2)$ for any 
clique $C=C_i$ or separator $C=S_i$ in 
(\ref{eqn_hiwd_1}).
Since $(S_y)_{CC}/(n-1)$ is an unbiased estimator of 
$\Sigma_{CC}$, this suggests taking
 $\Phi \propto S_y/(n-1)$.

\end{enumerate}

We assume in all cases that $\tau$ is uniform on 
the interval $\left[0, \Gamma \right] $ 
where $\Gamma$ is 
large, \eg $\Gamma=10^{10}$, and in the
equicorrelated case that $\rho$ 
is uniform on the open interval $(-1/(p-1),1)$.

\subsection{Prior for $g$}\label{prior:g}
We first define
notation for the \textit{edge indicators} 
of a graph $g$. \ Let%

\begin{equation}
e_{ij}=\left\{
\begin{array}
[c]{c}%
1\qquad\text{if }(i,j)\in E\\
0\qquad\text{otherwise \ }%
\end{array}
\right.  \label{def:edge}
\end{equation}
and let $e_{-ij}=\{e_{kl}:(k,l)\neq(i,j)\}.$ Note that 
any graph $g=(V,E)$ can be
unambiguously written as
$g=(e_{ij}, e_{-ij})$.

For a given graph $g=g(\Sigma)$, let the number of edges, or the \textit{size} of 
$g$, be given by
\begin{equation}
size(g)=\sum_{i<j} e_{ij}\label{def:size}
\end{equation}
\ie $size(g)$ is the number of nonzero elements in the strict upper triangle of
$\Omega$, and $size(g) \leq r=p(p-1)/2$.

Because of the theoretical and practical
difficulty in calculating for a given $p$ the exact number of 
decomposable graphs, or the number of graphs 
of a given size,
most of the literature for both decomposable and 
general models takes the prior for $g$ as uniform over all
the relevant graphs;
see, for example, \cite{Giudici_G99}, \cite{Dellaportas_F99},  \cite{Geiger_H02}, 
 \cite{Giudici_C03},   
\cite{Roverato02}, and \cite{Kayis_M}. 

Such a prior favours any class of graphs with many members over 
a class with few members, and favours middle sized graphs over
both very large and very small sized graphs.

Let $A_{p,k}$ denote the number of 
graphs of size $k$. 
We specify the prior for a graph $g$ hierarchically as follows. 
\[
p(g|size(g)=k)=\frac{1}{A_{p, k}},
\]
so that all graphs of a given size are equally likely. We now specify
the prior for the size of a graph. 
One choice is 
\[
p(size=k) \propto A_{p, k},
\]
which means that 
\[
p(g)=p(g, size(g))= p(g|size(g))p(size(g)) \propto constant,
\]
giving the uniform prior for $g$. A more flexible prior is of the form
\[
p(size=k| \psi) =  \binom{r}{k} \psi^k(1-\psi)^{r-k},
\]
where we interpret $\psi$ as the probability that 
any two vertices have a common edge. We could then put a prior 
on $\psi$. Suppose we take the prior for $\psi$ as a beta with 
parameters $a$ and $b$, \ie
\[
p(\psi) =  \frac{\psi^{a-1}(1-\psi)^{b-1}}{B(a,b)}.
\]
Then, 
\[
p(size=k) =  \binom{r}{k} \frac{B(a+k, r-k+b)}{B(a,b)}
\]
and 
\begin{align}
p(g) &=p(g|size(g))p(size(g)) \\
     &=\binom{r}{size(g)} \frac{B(a+size(g), r-size(g)+b)}{A_{p,size(g)}B(a,b)},
\end{align}
where $B(a,b)$ is the beta function.
We could now also put a prior on $a, b$. In our article we take 
$\psi$ uniform so that $a=b=1$, which means that 
\[
p(size=k)=\frac{1}{(r+1)} \: \text{ and }\:  p(g)=\frac{1}{(r+1) A_{p, k}} \ .
\]
That is, the size of each graph has equal probability, and the 
probability of a graph of size $k$ conditional on $size=k$ is uniform. 
However our framework is more flexible than this.

We call this the 
\textit{size based prior} for $g$ and 
compare results against those
using a uniform prior.

The size based prior makes it easier to discover sparse and 
full graphs 
when $n/p$ is small. The counts $A_{p,k}$ are 
not available in the literature. 
\sect\ref{sec:counting} gives results to calculate a subset of them
analytically, and shows how to 
evaluate the rest by simulation.

\section{Posterior inference and {M}arkov chain {Monte Carlo} sampling }\label{sec:sampling}
We use Markov Chain Monte Carlo (MCMC) simulation to obtain all 
posterior 
distributions.
The simulation involves the generation of the graphs $g$ and 
the parameters in $\Phi$ but not $\Sigma$ and $\mu$ 
which are integrated out. 
Thus, our sampling scheme is said to generate from 
\textit{reduced conditionals} and is 
therefore expected to be more efficient than the 
sampling schemes in \cite{Giudici_G99} and 
\cite{Wong_C_K03}
that generate $\Sigma$ as part 
of their sampling scheme. 

We note that iterates of $\mu$ and $\Sigma $ can also be generated 
in conjunction with
the simulation, but such iterates of $\mu$ and $\Sigma$ do not 
have any influence on the 
convergence properties or dependence structure of the 
reduced conditional simulation. 

The following theorems are useful in evaluating the 
conditional distributions
required in the simulations.  
The first theorem gives a conjugate
prior property of the HIW distribution. 

Let $S_y$ be defined by (\ref{eqn:Sy}) and define
\begin{equation}\label{def:Phi*} 
\Phi^{\ast}=\Phi+S_{y} \text{ and }
\delta^{\ast}=\delta+n-1 \ .  
\end{equation}

\begin{theorem}
\label{thm_bcsm}(Dawid and Lauritzen, 1993) 
For the Bayesian model specified
by (\ref{eqn_hiwd_1}) and (\ref{eqn_like_1})%
\[
\Sigma|y,\delta,\Phi,g\sim HIW(g,\delta^{\ast},\Phi^{\ast}).
\]
\end{theorem}
\begin{proof}
See Dawid and Lauritzen (1993) or \append\ref{proofs}.
\end{proof}

The next theorem gives an expression for the marginal likelihood. 

\begin{theorem}
\label{thm_marg_like}(Giudici, 1996) For the Bayesian model specified by
(\ref{eqn_hiwd_1}) and (\ref{eqn_like_1}),%
\begin{equation}
p(y|\delta,\Phi,g)=\left(  2\pi\right)  ^{-\left( (n-1)p/2\right)  }
\frac{h(g,\delta,\Phi)}{h(g,\delta^{\ast},\Phi^{\ast})}
\end{equation}
\end{theorem}
\begin{proof}
See Giudici (1996) or \append\ref{proofs} .
\end{proof}

\subsection{Sampling the graphs g \label{gen:g}}
We sample the graphs $g$ by generating the edge indicators one at a time,
conditional on $\delta, \Phi$ and $e_{-ij}=\{e_{kl}, (k,l) \neq (i,j), k < l \}$
using the following MH sampling scheme.

Using the notation of \sect\ref{sec:model},
let $g^c=(V, E^c)$ be the current graph of $\Sigma$, 
which is 
decomposable by construction with edge indicators 
$\{e^c_{kl}: \; 1 \leq k<l \leq p \}$.

We choose a pair $(i,j)$ at random and suppose that $g=(e_{ij}, e^c_{-ij})$
is decomposable for both $e_{ij}=0$ and $e_{ij}=1.$ 
We use the legal edge addition and deletion characterizations of 
\cite{Giudici_G99} and \cite{Frydenberg_L89} 
respectively to ensure this. Otherwise we choose a new
pair $(i,j)$. 

Set the proposal graph as $g^p$ (conditional
on $g^c$) as  
$g=(e^p_{ij}, e^c_{-ij})$ where  $e^p_{ij}=1-e^c_{ij}$. 
This means that the proposal density for $e_{ij}$ is 
$q_g(a|b, e^c_{-ij})$ where 
$a$ and $b$ are each either 
$ 0$ or $1$, and $q_g(a=1-b|b, e^c_{-ij})=1$. 

The MH acceptance probabilty for the proposal is 
\begin{equation}
\text{min}  \left\lbrace 1, \frac{p(y|g^p, \Phi, \delta) }{p(y|g^c, \Phi, \delta) } 
\frac{p(g^p)}{p(g^c)} 
\right\rbrace \label{mhratio:g}
\end{equation}
because $q_g(e^c_{ij}|e^p_{ij}, e^c_{-ij})/q_g(e^p_{ij}|e^c_{ij}, e^c_{-ij})=1$.
The ratio $p(g^p)/p(g^c)$ is known and the ratio of marginal likelihoods
\begin{equation}
\frac{p(y|g^p, \Phi, \delta) }{p(y|g^c, \Phi, \delta)}=\frac{h(g^p,\delta,\Phi)}{h(g^c,\delta,\Phi)}
\frac{h(g^c,\delta^{\ast},\Phi^{\ast})}{h(g^p,\delta^{\ast},\Phi^{\ast})}.\label{like_ratio:g}
\end{equation}
A simple expression for (\ref{like_ratio:g}) is derived in \append\ref{sec:theory}. 

\subsection{Generating the parameters in $\Phi$}
In all cases of \sect\ref{prior:phi} we generate $\tau$
using a random walk MH method 
\[
\operatorname*{log}(\tau^p)=\operatorname*{log}(\tau^c) + 
\xi_{\tau} \text{ }, \quad \xi_{\tau} \sim N(0, \sigma^2_{\tau}),
\]
which has acceptance probability
\begin{equation}
\text{min}  \left\lbrace 1, \frac{p(y|g, \tau^p, \rho) }{p(y|g, \tau^c, \rho)} 
\frac{p(\tau^p)}{p(\tau^c)} 
\right\rbrace 
\end{equation}
as the proposal densities cancel out.
In the equicorrelated case, the parameter $\rho$ is generated similarly to $\tau$ 
by a random walk
MH method 
\[
\rho^p=\rho^c + \xi_{\rho} \text{ }, \quad \xi_{\rho} \sim N(0, \sigma^2_{\rho}).
\]

The choice of the variances $\sigma^2_{\tau}, \sigma^2_{\rho}$ 
is sensitive to 
$p$, and was fine tuned to attain 
acceptance probabilities of around $25 \% $ 
according to the acceptance rate of the proposals.
For the case $p=17$ reported in this paper, such an acceptance probability
resulted from using $\sigma^2_{\tau}=1/10$ and  $\sigma^2_{\rho}=1/20$.

%
%
\subsection{Generating $\Sigma, \Omega$ and $\mu$\label{gen:sigma_mu}}
Although $\mu, \Sigma$ and $\Omega$ are not generated in the MCMC
simulation, it 
is often necessary to estimate functionals of $\mu, \Sigma$ and $\Omega$.
Such functionals can be estimated by sampling from the posterior 
distribution of $\Sigma, \Omega$ and $\mu$. Conditional on
$(g, \delta, \Phi)$ it follows
from
Theorem~\ref{thm_bcsm} that $p(\Sigma|y,g, \delta, \Phi)$ is HIW
 $(\delta +n-1, \Phi + S_y)$ so that 
$\Sigma$ and $\Omega$ can be generated using Theorems 3 and 4 of \cite{Roverato00}.
It is straightforward to show that 
$p(\mu| y, \Sigma, g, \delta, \Phi)$ is $N(\overline{y}, \Sigma / n)$, 
and hence to generate $\mu$, giving
iterates 
$\{ \mu^{ \left[ j \right] }, \Sigma^{ \left[ j \right]},\Omega^{ \left[ j \right] },\: j \geq 1\}$
from the posterior distribution.

\subsection{Efficient estimation of $E(\Omega|y)$\label{estimates}}
The posterior mean of $\Omega$ is not only used as an estimator of 
$\Omega$, but also of $\Sigma$ because $E(\Omega|y)^{-1}$ 
is the 
Bayes estimator of $\Sigma$ for the $L_1$ loss function in
\sect\ref{sec:size_vs_unif}. 
One method of estimating $E(\Omega|y)$ is to use the histogram
estimator $J^{-1}\sum^J_{j=1}\Omega^{[j]}$. 
A statistically more efficient estimator is the 
mixture estimator 
$J^{-1} \sum^J_{j=1} E(\Omega| y, g^{[j]}, \delta^{[j]}, \Phi^{[j]}).$                                    

We now show how to efficiently compute $E(\Omega|y, g, \delta,\Phi)$ 
using the following notation from
\cite{Lauritzen96}. \ Suppose that $A$ is a $p\times p$ matrix and $S\subset V$.
Let $B=\left[  A_{SS}\right]  ^{V}$ be the $p\times p$ matrix
defined by%
\[
B_{ij}=\left\{
\begin{array}
[c]{c}%
A_{ij}\qquad\text{if }\{i,j\}\subset S\\
0\qquad\text{otherwise \ \ \ \ }%
\end{array}
\right.
\]

\begin{theorem}
\label{thm_mixture}Suppose that $\Omega|y \sim W(g,\delta^{\ast},\Phi^{\ast})$,
where $g$ is decomposable. Then, using the notation of this and
\sect\ref{sec:model},%
\begin{equation}
E\left(  \Omega|y,\delta,\Phi,g\right)  =\sum_{i=1}^{k}\left[  \left(
\delta^{\ast}+\left|  C_{i}\right| -1 \right)  \left(  \Phi_{C_{i}C_{i}}^{\ast
}\right)  ^{-1}\right]  ^{V}-\sum_{i=2}^{k}\left[  \left(  \delta^{\ast
}+\left|  S_{i}\right| -1 \right)  \left(  \Phi_{S_{i}S_{i}}^{\ast}\right)
^{-1}\right]  ^{V}. \label{eqn_mixture}%
\end{equation}
\end{theorem}

\begin{proof}
See \append\ref{proofs}.
\end{proof}

\section{Comparison of the size prior for a graph with the uniform prior}
\label{sec:size_vs_unif}
This section compares the prior based on the graph size with 
the uniform prior that is used in most previous articles.
Performance is in terms of a loss function and a simulation was carried 
out to numerically assess performance. We found that overall the size 
based prior for $g$ outperformed the uniform prior.

Our simulation considered the following
five graph types for $g$.
(a) $\Omega = I$, the identity matrix, representing the empty graph 
and a diagonal covariance matrix; 
(b) $\Omega$ tridiagonal, representing a sparse and decomposable graph 
(this is a chain graph with $p-1$ edges); 
(c) $\Omega$ an `extreme' full matrix 
(the correlation coefficients $\rho_{ij}$ of $\Omega^{-1}$ satisfy 
$|\rho_{ij}| >.30$), which is a complete graph;
(d) $\Omega$ corresponding to a 4-cycle 
on $p$ vertices representing a sparse but nondecomposable graph; and
(e) $\Omega$ corresponding to a $p-$cycle on $p$ vertices, 
again representing a sparse but 
nondecomposable graph. 
We note that the nondecomposable graphs in (d)
and (e) require the addition of extra edges when we estimate them by a mixture of 
decomposable graphs.
Furthermore, 
(e) is an extreme case of non-decomposability, 
as it requires a minimum of $p-3$ fill ins.
Conversely, the unchorded 4-cycle on $p$ nodes
requires the fewest number of fill ins, so was chosen as an indicator 
of performance for the sparsest nondecomposable case.

The simulation considered the three forms of $\Phi$ described in 
\sect\ref{prior:phi} and two sample sizes $n= 40$ and $n = 100$. 
We report results for  matrices of size $p = 17$, but 
similar results were obtained for matrices of other sizes. 

Let $\Sigma_T$ be the true value of $\Sigma$ and let $\Sigmahat$ 
be an estimator of $\Sigma_T$. 
We measure the performance of $\Sigmahat$
using the $L_1$ loss function
\begin{equation}
L_1(\Sigmahat, \Sigma_T) = 
              \trace ( \Sigmahat \Sigma_T^{-1}) - \log \det (  \Sigmahat \Sigma_T^{-1}) - p.
\end{equation}

This loss function is frequently used to compare estimates of the 
covariance matrix, \eg \cite{yang94}.
It is straightforward to show that $L_1 \geq 0 $ for all $\Sigmahat$ and $\Sigma_T$, 
and that it is only equal to 0 if $\Sigmahat = \Sigma_T$. 
It is also straightforward to show that for $y \sim N(0, \Sigma)$, 
\begin{align}
L_1(\Sigmahat, \Sigma_T) & = - \int p(y|\Sigmahat) \log \biggl (  \frac{ p(y|\Sigma_T) } { p(y|\Sigmahat ) } 
\biggr )  dy 
\end{align}
\ie $L_1$ is equivalent to a Kullback-Liebler distance between $p(y|\Sigma_T)$ and $p(y|\Sigmahat ) $ with respect to the density $p(y|\Sigmahat)$.
The Bayes estimator for $\Sigma$ for the $L_1$ loss function is 
$E(\Omega|y)^{-1}$, which can be computed as in \sect\ref{estimates}.

We use boxplots to compare replication by replication
the size based prior with the uniform prior in terms of the 
percentage {\em increase} in the loss function $L_1$
resulting from using the uniform prior compared to the size-based 
prior; \ie the boxplots are based on calculating
\[
100(L_1^{unif} - L_1^{size})/L_1^{size}
\]
for each replication, where $L_1^{unif}$ and 
$L_1^{size}$ are the values of $L_1(\Sigmahat, \Sigma_T)$
for the uniform and size based priors respectively.

The boxplots are 
based on 20 replications with each replication consisting 
of 2,000 burnin iterations and 20,000 sampling iterations. 
We ran the sampler for the case $p=17$ on $n=40$ and $100$ 
observations from five simulated data sets corresponding
to the five models (a)--(e) for $\Omega$.

\begin{figure}[h!]
\centering
\includegraphics[angle=0,width=0.95\textwidth,height=0.55\textheight]{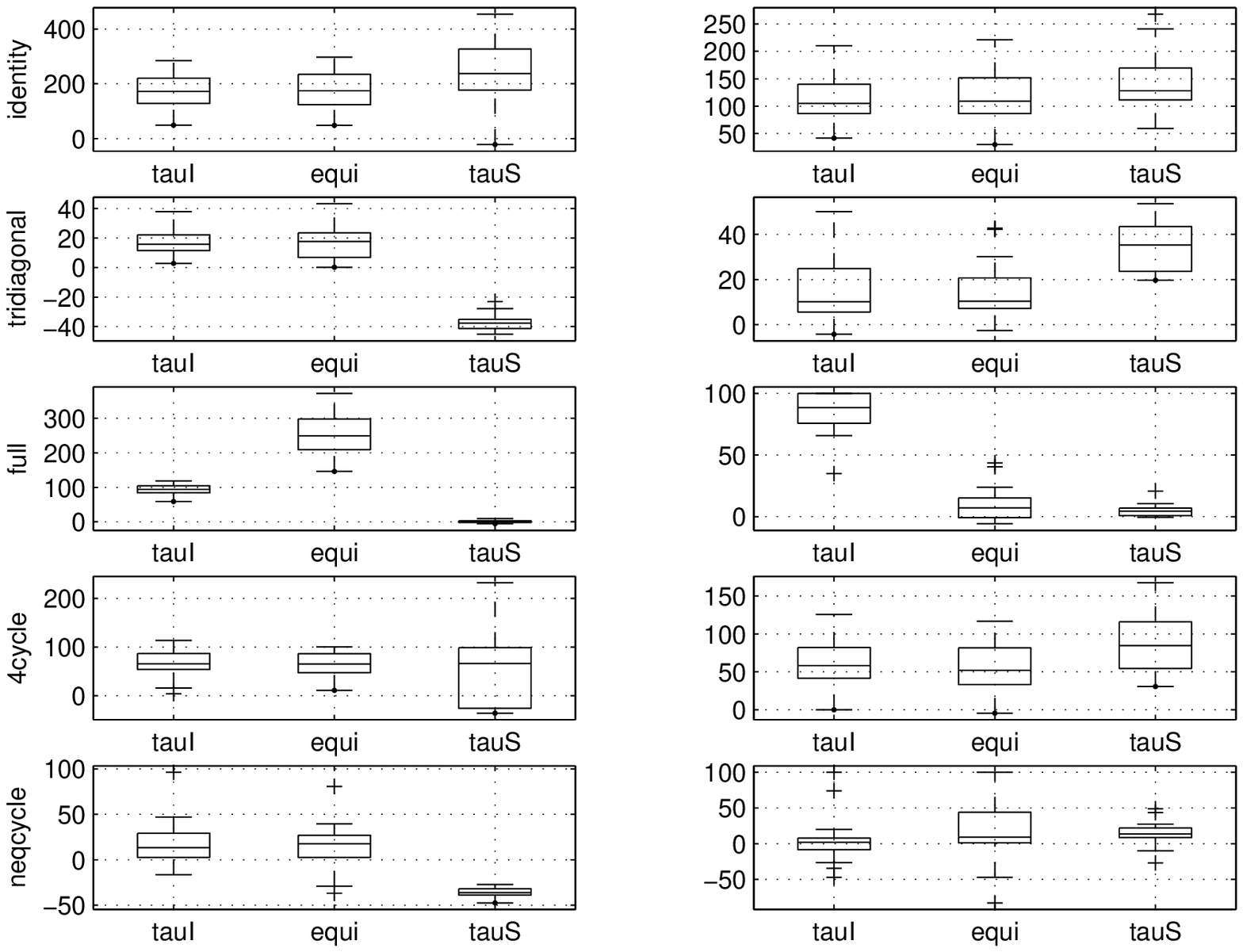}
\caption{Percentage increase in loss of uniform prior relative to the size prior
measured under $L_1$ loss. The left panels correspond to 
$n=40$ and the right panels to $n=100$. tauI, equi and tauS correspond 
to $\Phi=\tau I$, $\Phi$ equicorrelated and $\Phi=\tau S_y /(n-1)$.}
\label{fig:box17_L1_rel_incr}
\end{figure}

\fig\ref{fig:box17_L1_rel_incr}
presents the results for $p=17$.
The plots show that for 
$\Phi=\tau I$ and $\Phi$
equicorrelated, the size prior is at least as good, 
and often much better than, the uniform prior. For 
$\Phi=\tau S_y /(n-1)$, the comparison between the
size prior and the uniform prior is inconclusive for $n=40$, but
for $n=100$ the size prior is at least as good as, and often better 
than the uniform prior.
We conclude that the 
size based prior outperforms 
the uniform prior. 

We also compared the performance of the three forms of $\Phi$ 
for the uniform and size priors and found that overall
the equicorrelated form of $\Phi$
using the size based prior for the graph performed best, 
and it is this combination that we use for 
the rest of the paper.

\section{Evaluating the size based prior}\label{sec:counting}
To use the size based prior for graphs on $p$ vertices, we need the 
set of numbers $\{A_{p,k}: \: k=0, \dots, r\}$ where
$A_{p,k}$ is the number of decomposable graphs of size 
$k$ on $p$ vertices, and $r = \binom{p}{2}$ is the maximum graph size. 
These numbers are not in the literature, 
nor is there a general method available for computing them. 
In this section we present some exact values of $A_{p,k}$
as well as a simulation method that can estimate 
the $A_{p,k}$ as precisely as necessary. 

Let $B_{p,k}$ be the number of 
connected decomposable graphs of size $k$ on $p$ vertices.  
Equations~(3) and (4) of \cite{Castelo_W01} give recurrences
to calculate 
$A_{p,k}$ from the $B_{p,k}$ analytically, 
and the information
to calculate all $B_{p,k}$ analytically is implicit in
\cite{Wormald85}. For $p \leq 8$, \cite{Wormald85} gives the $B_{p,k}$
from which we computed the $A_{p,k}$
and these are reported in Table~\ref{table:Apk}.

\begin{table}
\caption{For each $p$, \: $2 \leq p \leq 8$ 
the table gives each $A_{p,k}, \: 0 \leq k \leq r$ and $A_{p}=\sum^r_{k=0} A_{p,k}$. 
The table also gives for each $p$ the percentage of graphs that are decomposable.
\label{table:Apk}}
\centering
\fbox{%
\begin{tabular}[c]{c|rrrrrrr}
\hline
k & 2 & 3 & 4 & 5 & 6 & 7 & 8\\
\hline \hline
0 & 1 & 1 & 1 & 1 & 1 & 1 & 1\\
1 & 1 & 3 & 6 & 10 & 15 & 21 & 28\\
2 & & 3 & 15 & 45 & 105 & 210 & 378\\
3 & & 1 & 20 & 120 & 455 & 1330 & 3276\\
4 & & & 12 & 195 & 1320 & 5880 & 20265\\
5 & & & 6 & 180 & 2526 & 18522 & 92988\\
6 & & & 1 & 140 & 3085 & 40647 & 315574\\
7 & & & & 90 & 3255 & 60795 & 770064\\ 
8 & & & & 30 & 3000 & 79170 & 1357818\\ 
9 & & & & 10 & 2235 & 92785 & 2078300\\ 
10 & & & & 1 & 1206 & 94521 & 2892176\\ 
11 & & & & & 615 & 81417 & 3621576\\ 
12 & & & & & 260 & 58485 & 4016439  \\ 
13 & & & & & 60 & 40110 & 3916724\\ 
14 & & & & & 15 & 24255 & 3432660  \\ 
15 & & & & & 1 & 12222 & 2855748\\ 
16 & & & & & & 4872 & 2185484\\ 
17 & & & & & & 1890 & 1488984 \\ 
18 & & & & & & 595 & 902944\\ 
19 & & & & & & 105 & 493220\\ 
20 & & & & & & 21 & 258468 \\ 
21 & & & & & & 1 & 118504\\ 
22 & & & & & &   & 46046\\ 
23 & & & & & &   & 14868\\ 
24 & & & & & &   & 4690\\ 
25 & & & & & &   & 1176\\ 
26 & & & & & &   & 168\\ 
27 & & & & & &   & 28\\ 
28 & & & & & &   & 1\\ 
\hline \hline
$\sum^r_{k=0} A_{p,k}$ & 2 & 8 & 61 & 822 & 18,154 & 617,675 & 30,888,596 \\
\hline
\% decomposable & 100\% & 100\% & 95\% & 80\% & 55\% & 29\%  & 12\%\\ 
\hline
\end{tabular}}
\end{table}

However, Wormald's (1985) analytic approach for obtaining the 
$B_{p,k}$ is likely to be computationally intractable for 
$p > 25$ (private correspondence with Wormald) and even for 
$8 < p \leq 25$ obtaining the $B_{p,k}$
would take
weeks on realistically sized computers. Furthermore, analytically
deriving the $A_{p,k}$ from the $B_{p,k}$ is computationally 
feasible only for small $p$. Because of these difficulties we
propose a simulation methodology to estimate the $A_{p,k}$ for
all $p$. 

\subsection{Methodology}\label{sec:method}
We begin with some exact results which can be used to calculate
$\{A_{p,k}: \: k \leq 5 \text{ and } r-2 \leq k \leq r\}$  
analytically for any $p$.
Let $F_{p,k}$ denote the number of nondecomposable graphs 
having $p$ vertices 
and $k$ edges.

\begin{lemma}\label{lemma:Apk_trivial} 
\begin{enumerate}
\item $ A_{p,k}=\binom{r}{k} - F_{p,k}$.
\item $  F_{p,0}=F_{p,1} = F_{p,r} =0, \: p \geq 0.$
\item $ F_{p,2}=F_{p, r-1}=0 , \: p \geq 2.$
\item $ F_{p,3}=0, \: p\geq 3.$
\end{enumerate}
\end{lemma}
\begin{proof}
The proof is obvious.
\end{proof}

\begin{lemma}\label{lemma:Apk_not_trivial} 
\begin{enumerate}
\item For $p \geq 4$, \quad $F_{p,4}= \binom{p}{4} \times 3. $
\item For $p \geq 4$, \quad $F_{p,r -2}=F_{p,4}.$
\item For $p \geq 5$, \quad 
$F_{p,5}=\binom{p}{5} \times 12 + \binom{p}{4}\times 3 \times ( r -6 ). $
\end{enumerate}
\end{lemma}
\begin{proof}
See \append\ref{proofs}.
\end{proof}

We now show how to estimate the 
$\{A_{p,k}: \: 6 \leq k \leq r-3 \}$ for all $p$.
Our approach is to run a separate simulation to estimate each $A_{p,k}$
for $6 \leq k \le r-3$.
The simulations are done in ascending order of $k$, i.e. $k = 6, \ldots, r-3$,
and the simulation to estimate a particular $A_{p,k}$ is restricted to graphs
of size $\leq k$ and uses the estimates $ \widehat{A}_{p,j}$ of $A_{p,j}$
for $j=6, \cdots, k-1$ that have been calculated in previous simulations.

We now describe the details of the simulation to estimate a particular $A_{p,k}$.
Let $\phi_{p,k}$ be the initial estimate of $A_{p,k}$ given by
\begin{equation}\label{eqn:phi_pk}
\phi_{p,k}=\widetilde{\alpha}_{p,k} \frac{\widehat{A}^2_{p,k-1}}
                                 {\widehat{A}_{p,k-2}}
\end{equation}
with $ \widetilde{\alpha}_{p,k} $ chosen in the range $(0.5,1)$. 
To justify this choice of $\phi_{p,k}$, we note that 
we have found empirically that $\log A_{p,k}$ is approximately 
a negative quadratic 
(see figures \ref{fig:A8k_results} and \ref{fig:A34k_results})
so that 
$\log A_{p,k} -2\log A_{p,k-1} + \log A_{p,k-2} \leq 0$, and hence 
\begin{equation*}
\alpha_{p,k}=\frac{A_{p,k} / A_{p,k-1}  }{A_{p,k-1} / A_{p,k-2}}
                          \leq 1.
\end{equation*}

We have also found empirically that $\alpha_{p,k}$ 
is likely to exceed 0.5.

As \begin{equation*}
A_{p,k}=\alpha_{p,k} \frac{A^2_{p,k-1}}
                                 {A_{p,k-2}}
\end{equation*}
the above discussion suggests the choice of 
$\phi_{p,k}$ in (\ref{eqn:phi_pk}).
Further details on the choice of
$\widetilde{\alpha}_{p,k}$ can be obtained from the authors.

We use Lemmas~\ref{lemma:Apk_trivial} and \ref{lemma:Apk_not_trivial},
the estimates $ \widehat{A}_{p,j}$ of $A_{p,j}$
for $j=6, \cdots, k-1$ that have been calculated in previous simulations,
and the initial estimate $\phi_{p,k}$ of $A_{p,k}$ given above to define
the following probability distribution $p_e(g)$ on the graphs $g$ of size $\leq k$.
To simplify the notation we omit subscripts for $p$ and $k$ in $p_e(g)$.

\begin{equation}\label{eqn:p_e1}
p_e(g) \propto \left\{ \begin{array}{ll}
\frac{1}{A_{p, size(g)}} & \mbox{if } 0 \leq \mbox{size}(g) \leq 5 \\
\frac{1}{\widehat{A}_{p, size(g)}} & \mbox{if } 6 \leq \mbox{size}(g) \leq k-1 \\
\frac{1}{\phi_{p, k}} & \mbox{if } \mbox{size}(g)  = k
\end{array} \right.
\end{equation}
which implies that
\begin{align*} \label{p_e_ratio_size}
\frac{p_e(size=k)}{p_e(size \leq 5)} &= \frac{ A_{p,k} / \phi_{p,k}}  
                          {\sum_{j=0}^5 A_{p,j} / A_{p,j}} \\ 
         &= \frac{1}{6}A_{p,k} / \phi_{p,k}
\end{align*}
and hence
\begin{equation*}
A_{p,k} = 6 \phi_{p,k} \frac{p_e(size=k)}{p_e(size \leq 5)}.
\end{equation*}

By running the simulation described below based on $p_e(g)$ we can 
estimate the ratio $p_e(size=k) / p_e(size \leq 5)$ by their
relative frequencies and hence obtain an estimate of 
\begin{equation*}\label{eqn:Apk_est_used}
\widehat{A}_{p,k} = 6 \phi_{p,k} \frac{\hat{p}_e(size=k)}{\hat{p}_e(size \leq 5)},
\end{equation*}
where $\hat{p}_e(size=k)$ and $\hat{p}_e(size \leq 5)$ are
the empirical relative frequencies.

The simulation uses the following MCMC sampling scheme. 
As in \sect\ref{gen:g}, we generate the edge indicators one
at a time conditional on the other edge indicators.
Let $ g^c=(V, E^c)$ be the current graph with edge indicators given by 
$\{e_{kl}: \: (k,l) \in E^c \}$.
We select an edge $(i,j)$ at random.
If $g=(e_{ij},e^c_{-ij} )$ corresponds to a decomposable graph of
size $\leq k $ for both $e_{ij}=0$ and $e_{ij}=1$ 
then we proceed, where we again use the 
legal edge addition and deletion characterizations of 
\cite{Giudici_G99} and \cite{Frydenberg_L89} respectively to test this.
Otherwise we select a new edge.
If we proceed, then we propose a new graph $g^p=(1-e^c_{ij}, e^c_{-ij} )$
and accept this graph with probability
\begin{equation*} \label{mh:Apk}
\text{min} \left\lbrace  1, p_e(g^p) / p_e(g^c) \right\rbrace  
\end{equation*}
which is evaluated using (\ref{eqn:p_e1}).

We note that at each stage we can also re-estimate 
$A_{p,j}, \: j =6, \cdots, k-1$. 

\subsection{Results} \label{sec:Apk_results}
\begin{figure}[h!]
\centering
\includegraphics[angle=0,width=0.75\textwidth,height=0.25\textheight]{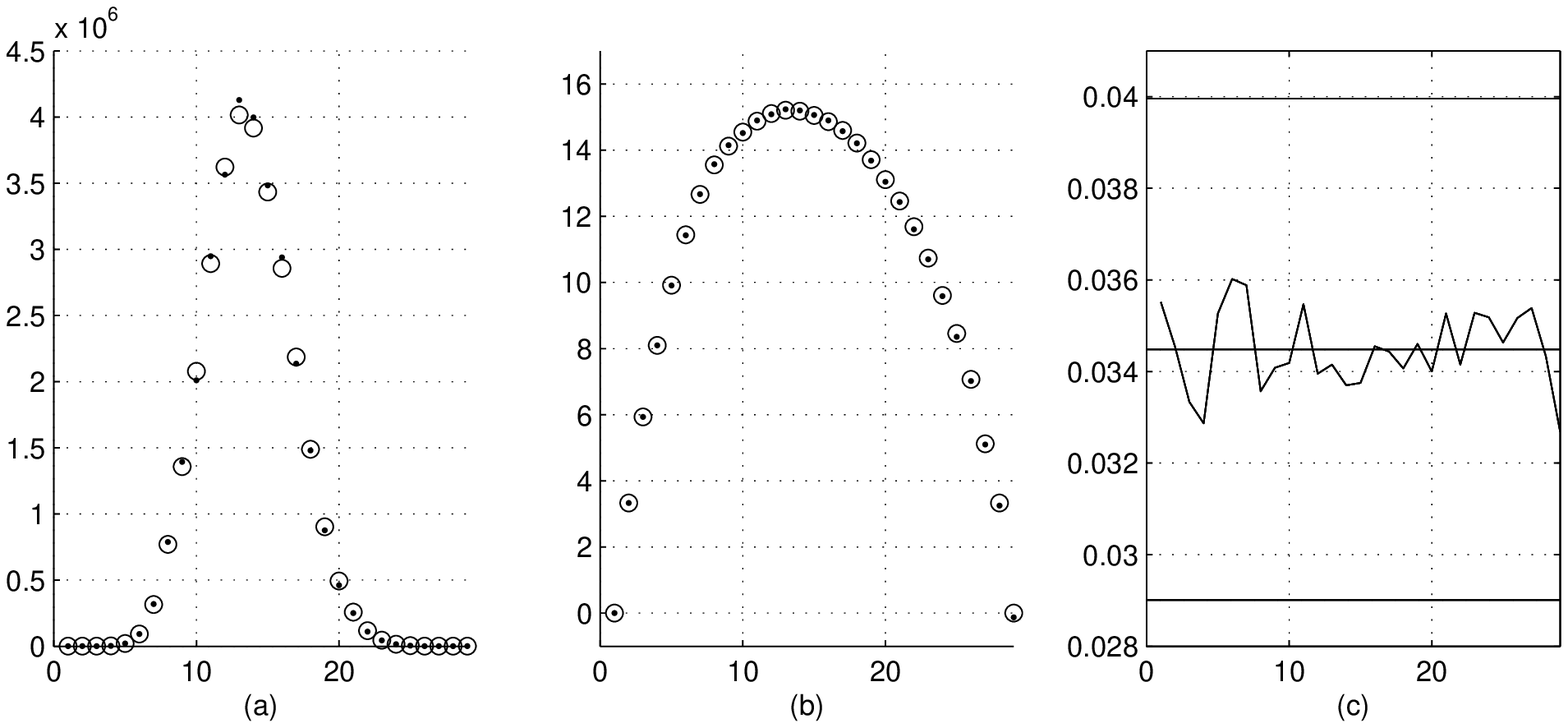}
\caption{Panel (a): Plot of true $A_{8,k}  (\cdot)$ and estimates $\widehat{A}_{8,k}$
 (open circles), \:$k=0, \cdots r$. 
Panel (b): Log scale of plot (a).
Panel (c): Plot of  $\hat{p}_e(size=k)$ together with 
their target value of $1/(r+1)$ (middle horizontal line) and $\pm 3$
approximate standard errors (outer horizontal lines).}
\label{fig:A8k_results}
\end{figure}

This section presents the estimates $\widehat{A}_{p,k}$
for $k=0 \cdots r$ and $p=8$ and $34$, and
provides a general method to check on the quality of these estimates.
Define the prior $p_e(g)$ on the decomposable graphs $g$ as
\begin{equation*}
p_e(g) \propto \left\{ \begin{array}{ll}
\frac{1}{A_{p, size(g)}} & \mbox{if } 0 \leq \mbox{size}(g) \leq 5 \mbox{ or } r-2 \leq \mbox{size}(g) \leq r \\
\frac{1}{\widehat{A}_{p, size(g)}} & \mbox{if } 6 \leq \mbox{size}(g) \leq r-3.
\end{array} \right.
\end{equation*}
The prior $p_e$ in this section is different to $p_e$ in 
\sect\ref{sec:method}. 
If the estimates $\widehat{A}_{p,k}$, 
$6 \leq k \leq r-3$ are precise,
then
$p_e(size=k)$ should be close to uniform
and hence close to the target value $1/(r+1)$.
An approximate lower bound for the standard error of the estimates of
$p_e(size=k)$ is $\sqrt{\pi(1-\pi)/J}$, where $\pi=1 / (r+1) $ and $J$ is the number of 
iterates used to compute $p_e(size=k)$.
Our simulations use a burnin period of 2,000 iterations and a sampling 
period of $N=10,000$ iterations.
\fig\ref{fig:A8k_results} plots the estimates 
$\widehat{A}_{p,k}$ for $p=8$ and the true values 
$A_{8,k}$, $k=0 \cdots r$
on both an absolute and logarithmic scale. 
\fig\ref{fig:A8k_results} also plots the estimates of $p_e(size=k)$ 
together 
with the target value $1/(r+1)$
and lower bounds for the $\pm 3$ standard error lines.

\begin{figure}[h!]
\centering
\includegraphics[angle=0,width=0.75\textwidth,height=0.25\textheight]{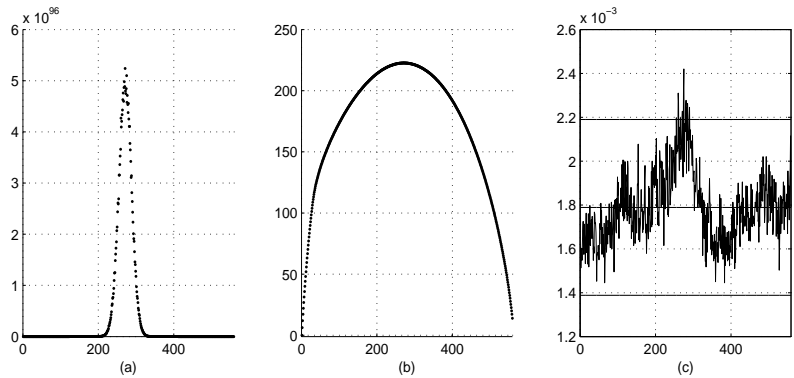}
\caption{Panel (a): Plot of estimates $\widehat{A}_{34,k}$ \:$k=0, \cdots r$. 
Panel (b): Log scale of plot (a).
Panel (c): Plot of  $\hat{p}_e(size=k)$ together with 
their target value of $1/(r+1)$ (middle horizontal line)
and $\pm 3$ approximate standard errors (outer horizontal lines).}
\label{fig:A34k_results}
\end{figure}

Figure~\ref{fig:A34k_results} has the same interpretation
as \fig\ref{fig:A8k_results} but is for $p=34$. The 
true values of $A_{34,k}$ are not plotted
as they are mostly unknown.

For $p=9, \cdots, 12 $ the totals $A_p=\sum_j A_{p,j}$ are known, 
but not the $A_{p,j}$. As a further check on results
we compared our estimated values of $\widehat{A}_p$ to $A_p$ 
and found that we were consistently within
$1\%$ of the truth.

\section{Comparsion of sampler efficiency to reversible jump approaches}\label{sec:us_vs_BGR}
This section compares the efficiency of our sampler to
the reversible jump approaches described in \cite{Brooks_G_R03}
using the six dimensional fowl bones dataset \citep{Whittaker90}.
To conform with the results given in \cite{Brooks_G_R03}
we use the equicorrelated form of $\Phi$ and the uniform prior
with a simulation run length of 1 million thinned to every 10th to give
$100 000$ generated graphs.

\begin{figure}[h!] \centering
\includegraphics[angle=0,width=0.7\textwidth,height=0.5\textheight]{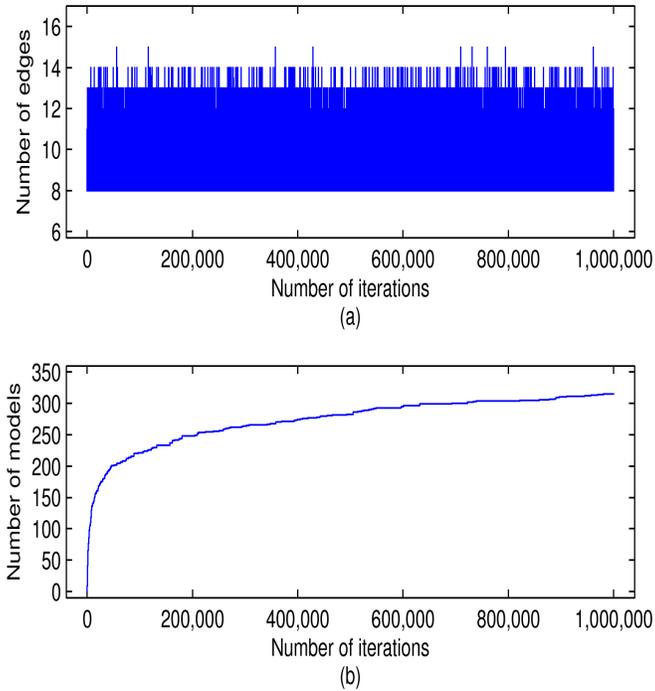}
\caption{Plots for the fowl bones dataset using the equicorrelated form of $\Phi$ and the uniform prior.
Note that there are \mbox{100 000} generated graphs, however, the horizontal axes run from 1 to 1 million iterates
to reflect the thinning process as described in the main text.
Panel (a) is the number of edges in each generated graph.
Panel (b) is the cumulative number of graphs visited during the simulation.
The total number visited is 315.}
\label{fig:fowlbones} \end{figure}

The plot of the number of edges in the generated graphs
given in Panel (a) of \fig\ref{fig:fowlbones} can be compared to \fig 2 of \cite{Brooks_G_R03}.
This plot shows that our sampler has much less dependence than the best performing approach in \cite{Brooks_G_R03}
which is the correlated AV method.
The plot of the cumulative number of graphs visited
given in Panel (b) of \fig\ref{fig:fowlbones} can be compared to \fig 3 of \cite{Brooks_G_R03}.
This plot shows that we visit 315 different graphs after \mbox{100 000} generated graphs in 1 million iterates.
This compares with 245 different graphs visited for the best performing method
in \cite{Brooks_G_R03} which is the correlated AV method.
Note that our sampler reaches the cumulative total of approximately 250 graphs
in the first \mbox{20 000} generated graphs which corresponds to the first \mbox{200 000} iterates.

A numerical comparison between the methods is given by
the effective sample size (ESS) \citep{Kass98}
for the thinned sample of \mbox{100 000} iterates of the number of edges
plotted in Panel (a) of \fig\ref{fig:fowlbones}.
The ESS for our method is \mbox{46 891}, which is over 30 times larger than
the best performing method in \cite{Brooks_G_R03} (the correlated AV method) which has
an ESS value of 1403.
Note that our ESS value is approximately 50\% of the maximum value of \mbox{100 000} for an independent sample.

\section{Physical measurements data}\label{sec:physical_p11responses_no_covariates}

In this section we illustrate our methods on a dataset consisting of 
the weight and various physical measurements described in \cite{Larner96}
on 22 male subjects aged 16 to 30.
The subjects were randomly chosen volunteers and were all in reasonably good health.
They were requested to slightly tense each muscle being measured to ensure measurement consistency. 
Apart from \textit{Mass}, all measurements are in cm.

The $p=11$ variables are indexed in the following order:
\small{
\begin{description}
\item[(1)] Mass: weight in kg, 
\item[(2)] Fore: maximum circumference of forearm,
\item[(3)] Bicep: maximum circumference of bicep,
\item[(4)] Chest: distance around chest directly under the armpits,
\item[(5)] Neck: distance around neck, approximately halfway up, 
\item[(6)] Shoulders: distance around shoulders, measured around the peak of the shoulder blades
\item[(7)] Waist: distance around waist, approximately trouser line,
\item[(8)] Height: from top of head to toe,
\item[(9)] Calf: maximum circumference of calf,
\item[(10)] Thigh: circumference of thigh, measured halfway between the knee and the top of the leg,
\item[(11)] Head: maximum circumference of head.
\end{description} }

\begin{figure}[h!] \centering \begin{tabular}{ccc} 
\includegraphics[angle=0,width=0.3\textwidth,height=0.22\textheight]{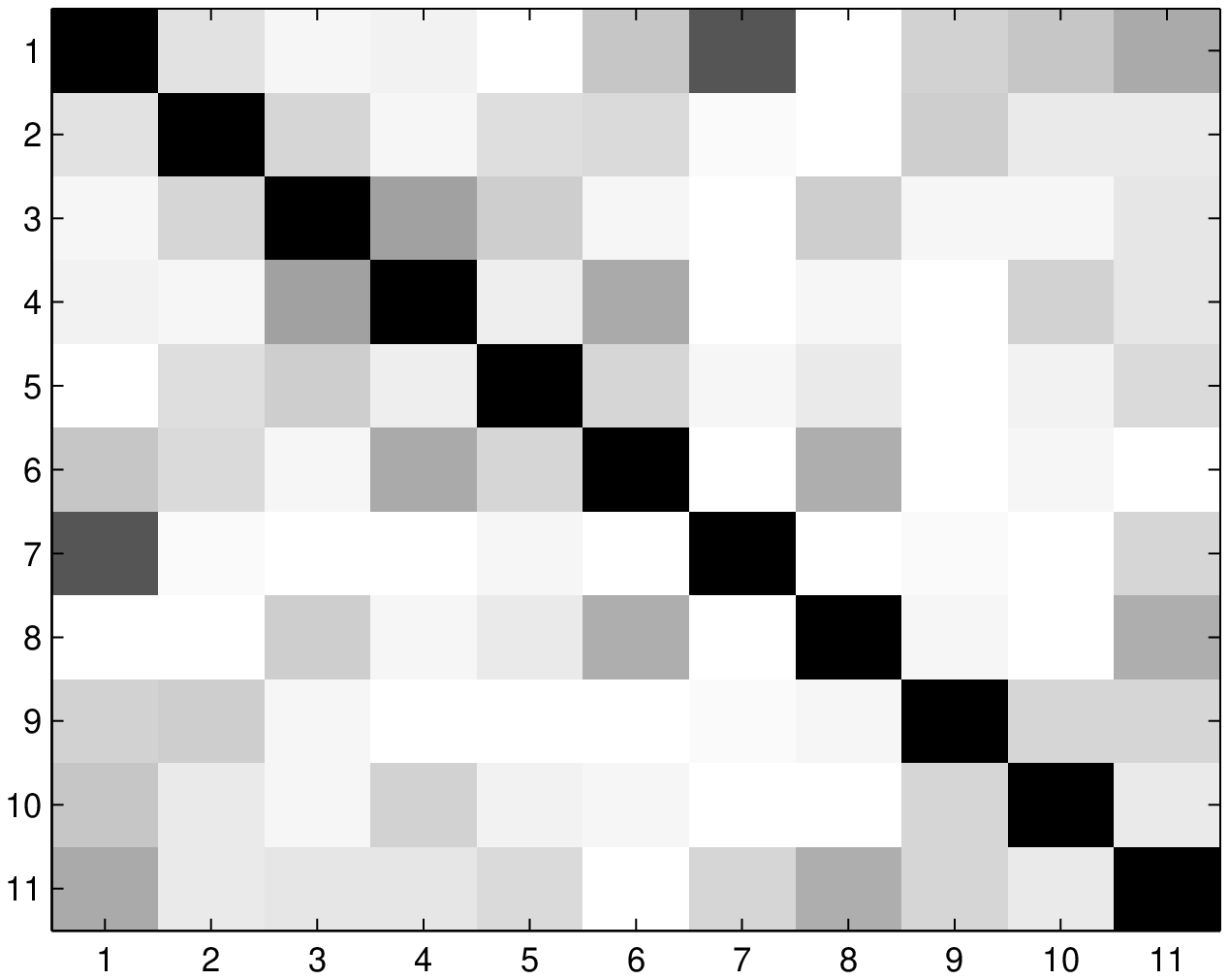}
&
\includegraphics[angle=0,width=0.3\textwidth,height=0.22\textheight]{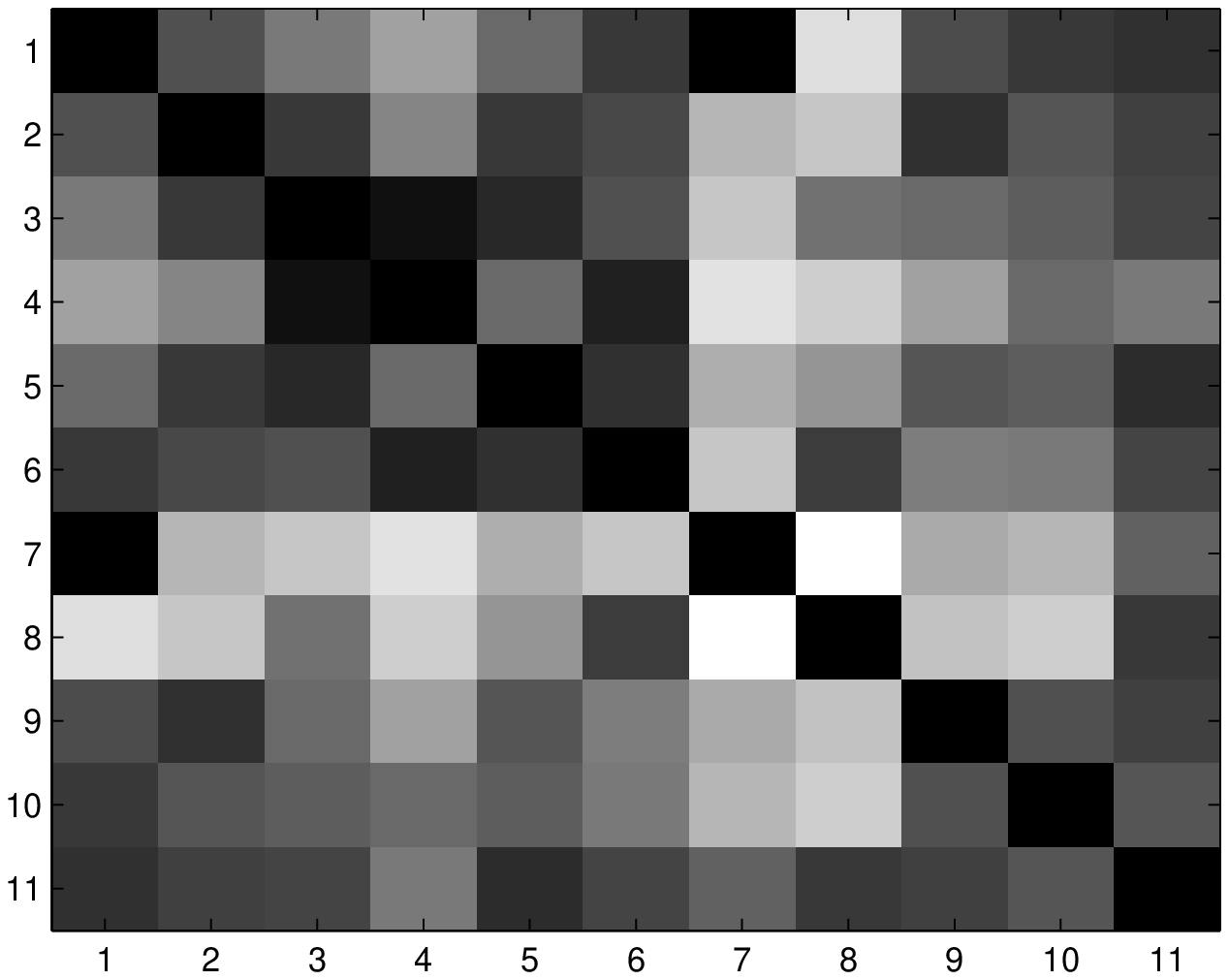}
&
\includegraphics[angle=0,width=0.3\textwidth,height=0.2\textheight]{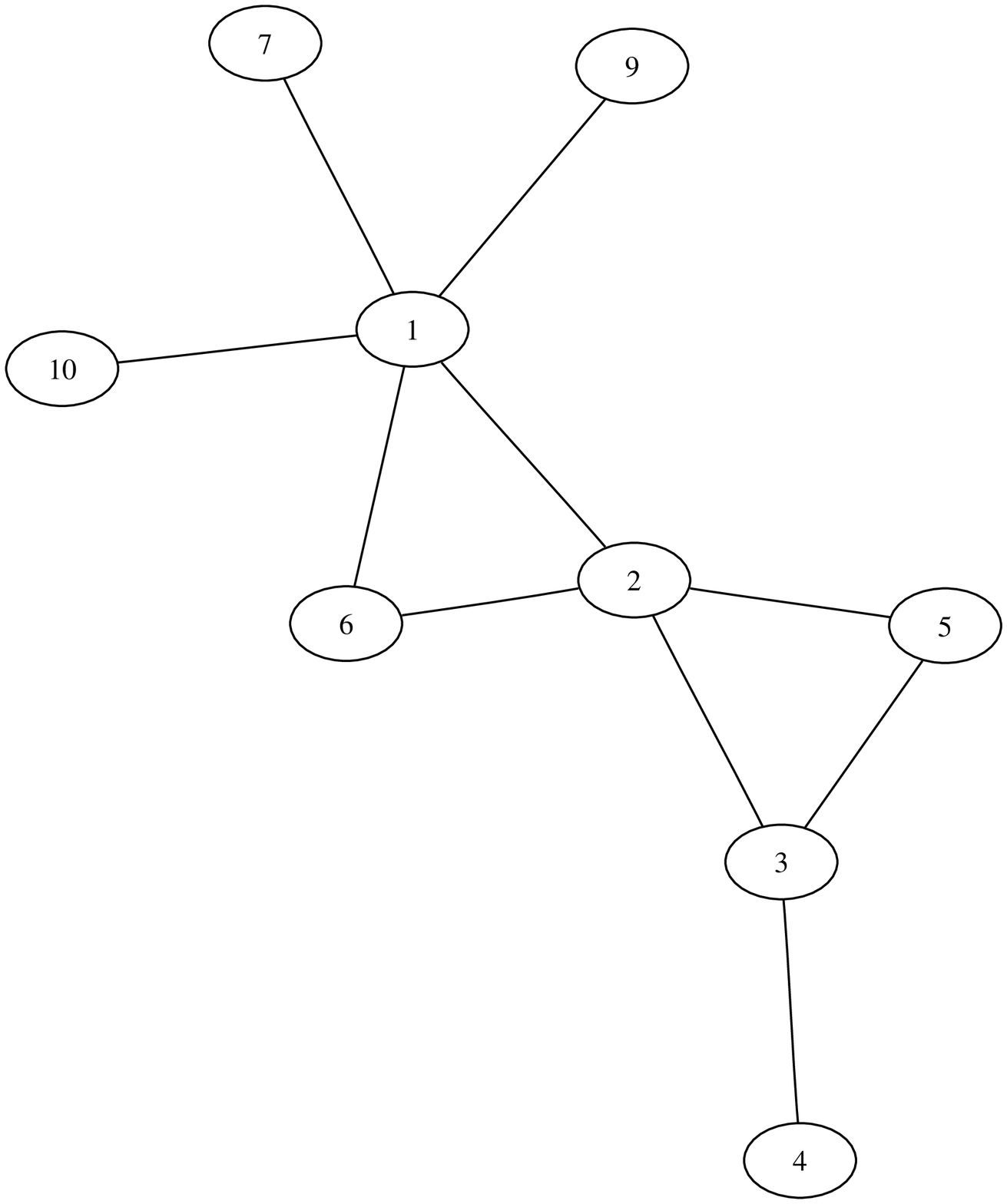} 
\\ 
\scriptsize{(a)} & \scriptsize{(b) } & \scriptsize{(c) }

\end{tabular} 
\caption{Plots for the physical measurements dataset using the equicorrelated form of $\Phi$
and the size based prior for the graph. 
Panel (a) is the image plot of the estimate of the partial correlation matrix.
Panel (b) is the image plot of $\Jhat$.
Panel (c) is the $70\%$ graph.}
\label{fig:physical_fix} \end{figure}

\normalsize
\fig\ref{fig:physical_fix} summarises the output using the equicorrelated form of $\Phi$
and the size based prior for the graph.
Panel (a) gives the estimate of the partial correlation matrix which equivalent to
$\Omegahat$ with the diagonal entries normalised to one.
Panel (b) gives the posterior probabilities of each edge being present.
Panel (c) gives the graph that results from applying a $70\%$ threshold to the values
in Panel (b).
Note that the procedure to obtain the graph in Panel (c) does not guarantee decomposability.

\section{Comparsion to the \cite{Wong_C_K03} covariance selection prior }\label{sec:us_vs_ed}
This section compares the performance of the prior in 
our article to the covariance selection prior of \cite{Wong_C_K03}, 
which does not assume that the graph of the covariance matrix is 
decomposable. 
Based on the results in \sect\ref{sec:size_vs_unif}, we 
use the equicorrelated form of $\Phi$ 
and the size based prior for the decomposable graphs.

The design of the simulation study is similar to that in 
\sect\ref{sec:size_vs_unif}. We use $L_1$ as the loss function,
$p=17$, two sample sizes $n= 40$ and $n = 100$, and four 
graphs for $\Omega$: identity, tridiagonal, 4-cycle and 
17-cycle. 

\begin{figure}[h!]
\centering
\includegraphics[angle=0,width=0.85\textwidth,height=0.25\textheight]
{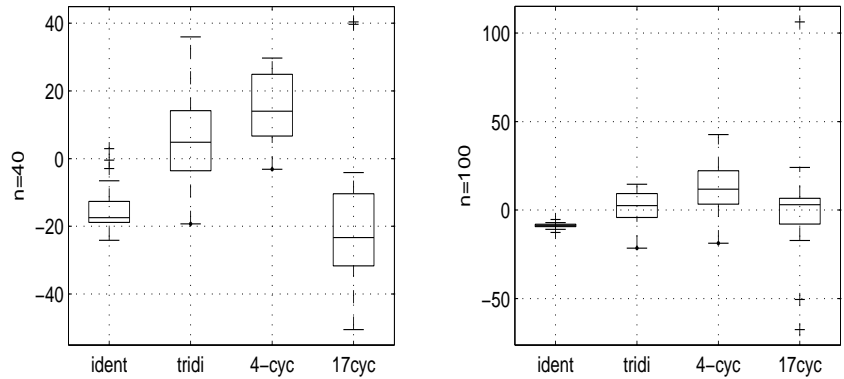}
\caption{Percentage increase in $L_1$ for $DCP$ over $NDP$.
The left panel is for $n=40$ and the right is for $n=100$.}
\label{fig:L1boxplots_us_vs_them}
\end{figure}

We refer to the decomposable prior as $DCP$ and the nondecomposable prior of 
\cite{Wong_C_K03} as $NDP$. \fig\ref{fig:L1boxplots_us_vs_them} 
reports boxplots of the percentage increase in $L_1$
of $DCP$ over $NDP$ for each iterate, \ie
\[
100(L_1^{DCP}-L_1^{NDP}) / L_1^{NDP}.
\]
\fig\ref{fig:L1boxplots_us_vs_them} shows that both priors 
perform similarly for decomposable graphs
and nondecomposable graphs, for both $n=40$ and $n=100$. 
These results and
others suggest that 
the prior based on decomposable graphs 
performs similarly to that of \cite{Wong_C_K03} when 
the graphs are relatively sparse.

Next we report autocorrelation plots for the iterates
of the elements of $\Omega$, when $p=5$ and the graph is full for 
both $DCP$ and $NDP$ when $n=40$. The simulation for $DCP$ 
uses a burnin of
50,000 iterations and a sampling of 50,000 iterations, and
$500,000$ burnin and $1$ million sampling iterations for $NDP$.

\begin{figure}[h!]
\centering
\includegraphics[angle=0,width=0.65\textwidth,height=0.45\textheight]
{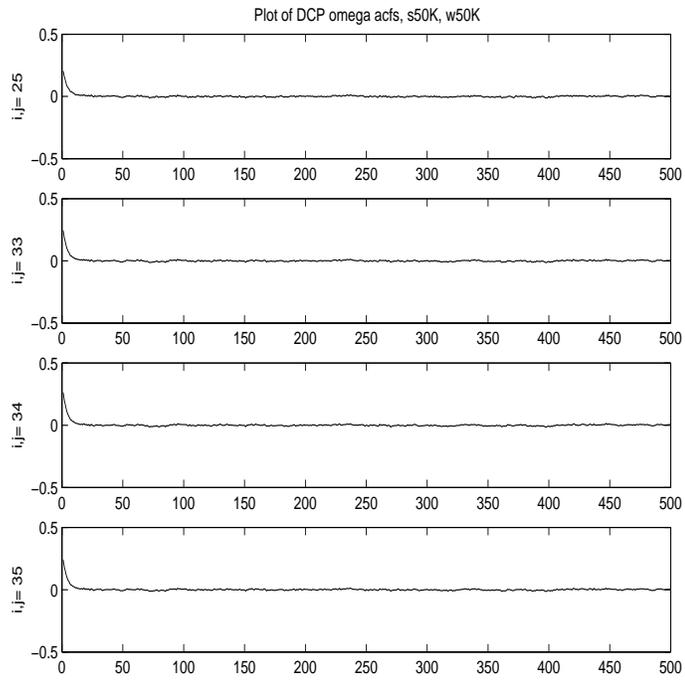}
\caption{Autocorrelations of the iterates of the $\Omega_{ij}$ 
in the $DCP$ case for a representative selection of $\Omega_{ij}$.}
\label{fig:acf_omega_helen}
\end{figure}

\begin{figure}[h!]
\centering
\includegraphics[angle=0,width=0.65\textwidth,height=0.45\textheight]
{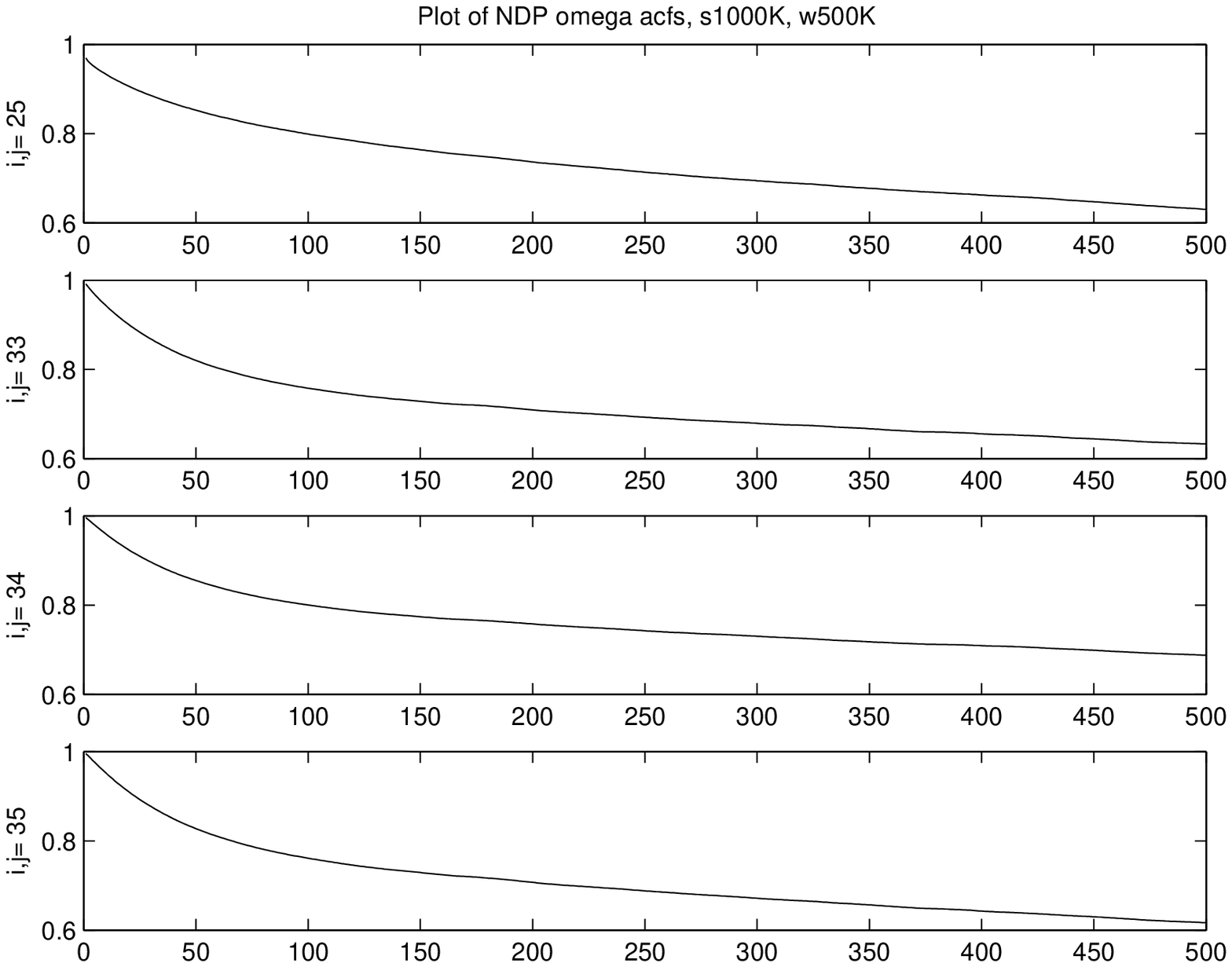}
\caption{Autocorrelations of the iterates of the $\Omega_{ij}$ 
in the $NDP$ case for a representative selection of $\Omega_{ij}$.}
\label{fig:acf_omega_ed}
\end{figure}

\begin{figure}[h!]
\centering
\includegraphics[angle=0,width=1.1\textwidth,height=0.30\textheight]
{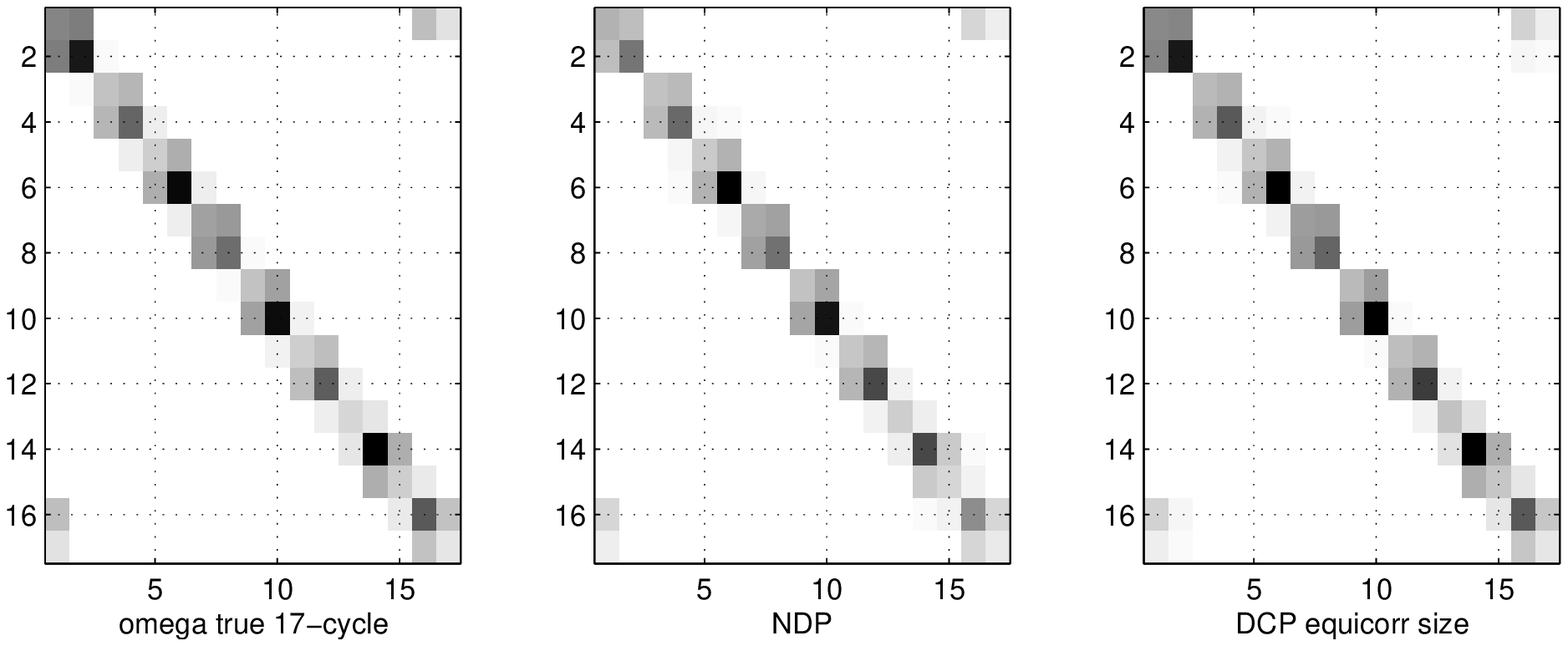}
\caption{True inverse covariance $\Omega$ and posterior mean 
estimates of  $\Omega$ for the $NDP$ estimator 
and the $DCP$ estimator for the 17-cycle case.}
\label{fig:omega40_pm_true_ed_us}
\end{figure}

Figures \ref{fig:acf_omega_helen} and \fig\ref{fig:acf_omega_ed}
are the autocorrelation plots for the $DCP$ and $NDP$ models
for a representative selection of $\Omega_{ij}$. 
The figures
show that the autocorrelations of the iterates of the 
$\Omega_{ij}$ decay rapidly to zero for the $DCP$ model, but are
far more dependent in the $NDP$ model. This difference in dependence
is due to the greater efficiency of the sampling scheme in the decomposable
case.
Grey scale plots of the true inverse covariance $\Omega$ 
and posterior mean 
estimates of  $\Omega$ for the $NDP$ estimator 
and the $DCP$ estimator for the 17-cycle case
indicated that $NDP$ and $DCP$
performed similarly in the simulations.
For brevity only the nondecomposable
17-cycle is presented as it represents a case
of high non-decomposability. Figure~\ref{fig:omega40_pm_true_ed_us} 
shows that even in this case,
the grey scales are very similar.

\clearpage
\appendix

\section{Proofs of results\label{proofs}}

\noindent\noindent\textbf{Proof} of Theorem \ref{thm_bcsm}\newline Roverato
(2000) shows that if $\Sigma\sim HIW(g,\delta,\Phi)$ and $\Omega=\Sigma^{-1}$
then%
\begin{equation}
p(\Omega|g,\delta,\Phi)\propto\left|  \Omega\right|  ^{\left(  \delta
-2\right) /2  }\operatorname*{etr}\left(  -\frac{1}{2}\Omega\Phi\right)
\label{eqn_gcwd_1}%
\end{equation} 
The result then follows from (\ref{eqn_like_1}) since%
\begin{align*}
p(\Omega|y, g, \delta,\Phi)  &  \propto p(y|\Omega)p(\Omega
|g, \delta,\Phi)\\
&  \propto\left|  \Omega\right|  ^{(n-1)/2}\operatorname*{etr}\left(
-\frac{1}{2}   \Omega S_{y}  \right)  |\Omega|^{\left(
\delta-2\right) /2 }\operatorname*{etr}\left(  -\frac{1}{2}\Omega
\Phi\right) \\
&  = |\Omega|^{\left(  n+\delta-3 \right) /2  }\operatorname*{etr}%
\left(  -\frac{1}{2}\Omega\left(  S_{y}+\Phi\right)  \right). 
\end{align*}
Note that the conjugate prior result for $\Omega$ does not require the graph
$g$ to be decomposable.

\noindent\textbf{Proof} of Theorem \ref{thm_marg_like}\newline First%
\[
p(Y|\delta,\Phi,g)=\frac{p(Y|\Sigma,\delta,\Phi,g)p(\Sigma|\delta,\Phi
,g)}{p(\Sigma|Y,\delta,\Phi,g)}.
\]
The result then follows from (\ref{eqn_hiwd_1}), (\ref{eqn_iwd_2}),
(\ref{eqn_like_1}) and \ Theorem \ref{thm_bcsm}.

\noindent\textbf{Proof} of Theorem \ref{thm_mixture}\newline From Equation
(5.23), Lemma 5.5 of \cite{Lauritzen96}
\[
\Omega=\sum_{i=1}^{k}\left[  \left(  \Sigma_{C_{i}C_{i}}\right)  ^{-1}\right]
^{V}-\sum_{i=2}^{k}\left[  \left(  \Sigma_{S_{i}S_{i}}\right)  ^{-1}\right]
^{V}%
\]
and hence
\[
E\left(  \Omega|Y,\delta,\Phi,g\right)  =\sum_{i=1}^{k}\left[  E\left(
\left(  \Sigma_{C_{i}C_{i}}\right)  ^{-1}|Y,\delta,\Phi,g\right)  \right]
^{V}-\sum_{i=2}^{k}\left[  E\left(  \left(  \Sigma_{S_{i}S_{i}}\right)
^{-1}|Y,\delta,\Phi,g\right)  \right]  ^{V}.
\]
Now $\Sigma|Y,\delta,\Phi,g\sim\operatorname*{HIW}\left(  \delta,\Phi^{\ast
},g^{\ast}\right)  $, so from Dawid and Lauritzen (1993), if $A$ is a complete
set in $g$ then $\left(  \Sigma_{AA}\right)  ^{-1}|Y,\delta,\Phi
,g\sim\operatorname*{Wishart}\left(  \delta^{\ast}+\left|  A\right|
-1,\Phi_{AA}^{\ast}\right)  $. The result then follows from the properties of
the Wishart distribution.

\noindent\noindent\textbf{Proof} of Lemma \ref{lemma:Apk_not_trivial}
\begin{enumerate}
\item For a nondecomposable graph to have 4 edges it must contain exactly one 
chordless 4-cycle and no other edges. There are $\binom{p}{4}$ possible 
choices for the 4 vertices, and for each choice of 4 vertices there are 
3 different chordless 4-cycles.

\item For a graph to be nondecomposable with $\binom{p}{2}-2$ edges it must contain 
exactly one 4 cycle and all other edges must be present. Then apply the proof of 
the above.

\item We can partition the nondecomposable graphs with 5 edges into 2 sets: \
(a) those with a chordless 5-cycle and no other edges, and 
(b) those with a chordless 4-cycle and an extra edge. 
For case (a) there are $\binom{p}{5}$ choices for the 5 vertices and for 
each choice there are $(5-1)!/2=12$ different chordless 5-cycles. 
For case (b) there are  $\binom{p}{4} \times 3$ choices for the chordless 
4-cycle, and for each choice of chordless 4-cycle 
there are $(\binom{p}{2} -6)$ choices for the extra vertex pair 
constituting the edge.
\end{enumerate}

\section{HIW results for Bayesian analysis using MCMC}
\label{sec:theory}
The following results derive an expression for (\ref{like_ratio:g})
that can
be evaluated efficiently. The first theorem gives 
some necessary graph theory.

Let $g=(V,E)$ be a
decomposable graph with edge indicators $\{ e_{ij}, i<j \leq p \}$. 
Assume the edge indicator $e_{ij}=1$ for $g$, and that
the graph $g^{\prime}=(V,E^{\prime})$ is decomposable and has
edge set $E^{\prime}$ 
as defined by indicators $\{e^{\prime}_{ij}=0, e_{-ij} \}$.

\begin{theorem}
\label{thm_single_clique}
Suppose that $g$ and $g^{\prime}$ are the
decomposable graphs defined above. Suppose that $C_{1},\ldots,C_{k}$ are the
cliques of $g$ ordered to form a perfect sequence and $S_{2},\ldots,S_{k}$ are
the corresponding separators. Then\newline (a) The edge $(i,j)$ is contained
in a single clique of $g$.\newline (b) If $(i,j)\in C_{q}$ then either
$i\notin S_{q}$ or $j\notin S_{q}$.\newline (c) If $j\notin S_{q}$ and
$C_{q_{1}}=C_{q}\backslash\{j\}$ and $C_{q_{2}}=C_{q}\backslash\{i\}$ then
$C_{1}$, $\ldots$, $C_{q-1}$, $C_{q_{1}}$, $C_{q_{2}}$, $C_{q+1}$, $\ldots$,
$C_{k}$ is a perfect sequence of complete sets in $g^{\prime}$ and has
separators $S_{2}$, $\ldots$, $S_{q-1}$, $S_{q_{1}}=S_{q}$, $S_{q_{2}}%
=C_{q}\backslash\{i,j\}$, $S_{q+1}$, $\ldots$, $S_{k}$.\newline (d) The
sequence $C_{1}$, $\ldots$, $C_{q-1}$, $C_{q_{1}}$, $C_{q_{2}}$, $C_{q+1}$,
$\ldots$, $C_{k}$ contains all the cliques of $g^{\prime}$.
\end{theorem}

\begin{proof}
Part (a) is Theorem 1 of \cite{Frydenberg_L89} 
.\newline
Parts (b) and (c) follow from part (a) and Lemma 2.20 of \cite{Lauritzen96}. \newline
To show part (d), suppose that $C^{\ast}$ is a clique of $g^{\prime}$.
Then $C^{\ast}$ is complete in $g$, so $C^{\ast}\subset C_{l}$
for some $l\in\{1,\ldots,k\}$.
If $C^{\ast}\subset C_{q}$ then part (b) implies that either $i\notin S_{q}$
or $j\notin S_{q}$. \ So either
$C^{\ast}\subset C_{q_{1}}$ or $C^{\ast}\subset C_{q_{2}}$. Hence $C^{\ast}$
is contained in at least one of $C_{1}$, $\ldots$, $C_{q-1}$, $C_{q_{1}}$,
$C_{q_{2}}$, $C_{q+1}$, $\ldots$, $C_{k}$. \ Part (c) shows that $C_{1}$,
$\ldots$, $C_{q-1}$, $C_{q_{1}}$, $C_{q_{2}}$, $C_{q+1}$, $\ldots$, $C_{k}$
are complete sets in $g^{\prime}$ and the result follows.

\end{proof}

The next lemma uses (\ref{eqn_hiwd_2}) and Theorem \ref{thm_single_clique} to
simplify (\ref{like_ratio:g}).

\begin{lemma}
\label{lemma_post_ratio}Suppose that $g$ and $g^{\prime}$ are the decomposable
graphs defined above. Then, using the notation of 
(\ref{def:Phi*}),  
 and 
Theorem~\ref{thm_single_clique} %
\begin{align}
\frac{h(g,\delta,\Phi)}{h(g^{\prime},\delta,\Phi)} &  \frac{h(g^{\prime
},\delta^{\ast},\Phi^{\ast})}{h(g,\delta^{\ast},\Phi^{\ast})}\nonumber\\
= &  \frac{\left|  \Phi_{DD|S_{q_{2}}}\right|  ^{\left(  \frac{\delta+\left|
S_{q_{2}}\right|  +1}{2}\right)  }\left|  \Phi_{ii|S_{q_{2}}}^{\ast}\right|
^{\left(  \frac{\delta^{\ast}+\left|  S_{q_{2}}\right|  }{2}\right)  }\left|
\Phi_{jj|S_{q_{2}}}^{\ast}\right|  ^{\left(  \frac{\delta^{\ast}+\left|
S_{q_{2}}\right|  }{2}\right)  }}{\left|  \Phi_{ii|S_{q_{2}}}\right|
^{\left(  \frac{\delta+\left|  S_{q_{2}}\right|  }{2}\right)  }\left|
\Phi_{jj|S_{q_{2}}}\right|  ^{\left(  \frac{\delta+\left|  S_{q_{2}}\right|
}{2}\right)  }\left|  \Phi_{DD|S_{q_{2}}}^{\ast}\right|  ^{\left(
\frac{\delta^{\ast}+\left|  S_{q_{2}}\right|  +1}{2}\right)  }}\times
\nonumber\\
&  \frac{\Gamma\left(  \frac{\delta+\left|  S_{q_{2}}\right|  }{2}\right)
\Gamma\left(  \frac{\delta^{\ast}+\left|  S_{q_{2}}\right|  +1}{2}\right)
}{\Gamma\left(  \frac{\delta+\left|  S_{q_{2}}\right|  +1}{2}\right)
\Gamma\left(  \frac{\delta^{\ast}+\left|  S_{q_{2}}\right|  }{2}\right)
},\label{eqn_like_ratio}%
\end{align}
where $D=\{i,j\}$, $\Phi_{DD|S_{q_{2}}}=\Phi_{DD}-\Phi_{DS_{q_{2}}}\left(
\Phi_{S_{q_{2}}S_{q_{2}}}\right)  ^{-1}\Phi_{S_{q_{2}}D}$, and $\Phi
_{ii|S_{q_{2}}}$, $\Phi_{jj|S_{q_{2}}}$, $\Phi_{DD|S_{q_{2}}}^{\ast}$,
$\Phi_{ii|S_{q_{2}}}^{\ast}$ and $\Phi_{jj|S_{q_{2}}}^{\ast}$ are defined similarly.
\end{lemma}

\begin{proof}
To obtain an expression for $h(g^{\prime},\delta,\Phi)$ we require the following technical
lemma based on Lemma 2.13 of \cite{Lauritzen96}.

\begin{lemma}
\label{lemma_thinning}Let $\widetilde{C}_{1}$, $\ldots$, $\widetilde
{C}_{\widetilde{k}\text{ }}$ be a perfect sequence with separators
$\widetilde{S}_{2}$, $\ldots$, $\widetilde{S}_{\widetilde{k}}$. \ Assume that
$\widetilde{C}_{t}\subset\widetilde{C}_{p}$ for some $t\neq p$ and that $p$ is
minimal with this property for fixed $t$. Then\newline (a) If $p<t$ then
$\widetilde{S}_{t}=\widetilde{C}_{t}$ and $\widetilde{C}_{1}$, $\ldots$,
$\widetilde{C}_{t-1}$, $\widetilde{C}_{t+1}$, $\ldots$, $\widetilde
{C}_{\widetilde{k}}$ is a perfect sequence with separators $\widetilde{S}_{2}%
$, $\ldots$, $\widetilde{S}_{t-1}$, $\widetilde{S}_{t+1}$, $\ldots$,
$\widetilde{S}_{\widetilde{k}\text{ }}$\newline (b)If $p>t$ then
$\widetilde{S}_{p}=\widetilde{C}_{t}$ and $\widetilde{C}_{1}$, $\ldots$,
$\widetilde{C}_{t-1}$, $\widetilde{C}_{p}$, $\widetilde{C}_{t+1}$, $\ldots$,
$\widetilde{C}_{p-1}$, $\widetilde{C}_{p+1}$, $\widetilde{C}_{\widetilde
{k}\text{ }}$ is a perfect sequence with separators $\widetilde{S}_{2}$,
$\ldots$, $\widetilde{S}_{t-1}$, $\widetilde{S}_{t}$, $\widetilde{S}_{t+1}$,
$\ldots$, $\widetilde{S}_{p-1}$, $\widetilde{S}_{p+1}$, $\widetilde
{S}_{\widetilde{k}\text{ }}$
\end{lemma}

\noindent\textbf{Proof} of Lemma \ref{lemma_thinning}. \ See Lemma 2.13 of
Lauritzen and its proof.\newline From Lemma \ref{lemma_thinning}, a perfect
sequence of complete sets $\widetilde{C}_{1}$, $\ldots$, $\widetilde
{C}_{\widetilde{k}\text{ }}$ containing the cliques of $g^{\prime}$ can be
thinned by removing complete sets that are not cliques and reordering the
sequence. \ From Lemma \ref{lemma_thinning}, the right-hand side of
(\ref{eqn_hiwd_2}) is invariant to this thinning process. \ Successive
application of the thinning process gives a perfect sequence consisting of the
cliques of $g^{\prime}$.

From (\ref{eqn_hiwd_2}), Theorem \ref{thm_single_clique} and Lemma
\ref{lemma_thinning}%
\begin{align}
&  h(g^{\prime},\delta,\Phi)\nonumber\\
&  =\frac{%
{\displaystyle\prod_{i=1,\ldots q-1,q_{1},q_{2},q+1,\ldots,k}}
\left[  \left|  \frac{\Phi_{C_{i}C_{i}}}{2}\right|  ^{\left(  \frac
{\delta+\left|  C_{i}\right|  -1}{2}\right)  }\Gamma_{\left|  C_{i}\right|
}\left(  \frac{\delta+\left|  C_{i}\right|  -1}{2}\right)  ^{-1}\right]  }{%
{\displaystyle\prod_{i=2,\ldots q-1,q_{1},q_{2},q+1,\ldots,k}}
\left[  \left|  \frac{\Phi_{S_{i}S_{i}}}{2}\right|  ^{\left(  \frac
{\delta+\left|  S_{i}\right|  -1}{2}\right)  }\Gamma_{\left|  S_{i}\right|
}\left(  \frac{\delta+\left|  S_{i}\right|  -1}{2}\right)  ^{-1}\right]  }.
\label{eqn_hiwd_3}%
\end{align}
Now consider the ratio $h(g,\delta,\Phi)/h(g^{\prime},\delta,\Phi)$.
Simplifying the expressions from (\ref{eqn_hiwd_2}) and (\ref{eqn_hiwd_3})
gives%
\begin{equation}
\frac{h(g,\delta,\Phi)}{h(g^{\prime},\delta,\Phi)}=\frac{\left|  \Phi
_{C_{q}C_{q}}\right|  ^{\left(  \frac{\delta+\left|  S_{q_{2}}\right|  +1}%
{2}\right)  }\left|  \Phi_{S_{q}S_{q}}\right|  ^{\left(  \frac{\delta+\left|
S_{q_{2}}\right|  -1}{2}\right)  }\Gamma\left(  \frac{\delta+\left|  S_{q_{2}%
}\right|  }{2}\right)  }{\left|  \Phi_{C_{q_{1}}C_{q_{1}}}\right|  ^{\left(
\frac{\delta+\left|  S_{q_{2}}\right|  }{2}\right)  }\left|  \Phi_{C_{q_{2}%
}C_{q_{2}}}\right|  ^{\left(  \frac{\delta+\left|  S_{q_{2}}\right|  }%
{2}\right)  }\Gamma\left(  \frac{\delta+\left|  S_{q_{2}}\right|  +1}%
{2}\right)  2\sqrt{\pi}}. \label{eqn_hratio_1}%
\end{equation}
Substituting%
\begin{align*}
\left|  \Phi_{C_{q}C_{q}}\right|   &  =\left|  \Phi_{DD|S_{q_{2}}}\right|
\left|  \Phi_{S_{q_{2}}}\right| \\
\left|  \Phi_{C_{q_{1}}C_{q_{1}}}\right|   &  =\left|  \Phi_{ii|S_{q_{2}}%
}\right|  \left|  \Phi_{S_{q_{2}}}\right| \\
\left|  \Phi_{C_{q_{2}}C_{q_{2}}}\right|   &  =\left|  \Phi_{ii|S_{q_{2}}%
}\right|  \left|  \Phi_{S_{q_{2}}}\right|
\end{align*}
into (\ref{eqn_hratio_1}) gives%
\begin{equation}
\frac{h(g,\delta,\Phi)}{h(g^{\prime},\delta,\Phi)}=\frac{\left|
\Phi_{DD|S_{q_{2}}}\right|  ^{\left(  \frac{\delta+\left|  S_{q_{2}}\right|
+1}{2}\right)  }\Gamma\left(  \frac{\delta+\left|  S_{q_{2}}\right|  }%
{2}\right)  }{\left|  \Phi_{ii|S_{q_{2}}}\right|  ^{\left(  \frac
{\delta+\left|  S_{q_{2}}\right|  }{2}\right)  }\left|  \Phi_{jj|S_{q_{2}}%
}\right|  ^{\left(  \frac{\delta+\left|  S_{q_{2}}\right|  }{2}\right)
}\Gamma\left(  \frac{\delta+\left|  S_{q_{2}}\right|  +1}{2}\right)
2\sqrt{\pi}}.\nonumber
\end{equation}
A similar expression can be derived for the ratio $h(g,\delta^{\ast}%
,\Phi^{\ast})/h(g^{\prime},\delta^{\ast},\Phi^{\ast})$ and the result follows.

\end{proof}

The following lemma gives an efficient method for evaluating the terms in
(\ref{eqn_like_ratio}) using Cholesky decompositions.

\begin{lemma}
\label{lemma_chol} 
Using the notation of Theorem \ref{thm_single_clique} and
Lemma \ref{lemma_post_ratio}, suppose that the matrix $A_{C_{q}C_{q}}>0$ is
partitioned as%
\[
A_{C_{q}C_{q}}=\left(
\begin{array}
[c]{cc}%
A_{S_{q_{2}}S_{q_{2}}} & A_{S_{q_{2}}D}\\
A_{DS_{q_{2}}} & A_{DD}%
\end{array}
\right)
\]
and has Cholesky decomposition $A_{C_{q}C_{q}}=LL^{\prime}$ where%
\[
L=\left(
\begin{array}
[c]{cc}%
L_{S_{q_{2}}S_{q_{2}}} & 0\\
L_{DS_{q_{2}}} & L_{DD}%
\end{array}
\right)
\]
and%
\[
L_{DD}=\left(
\begin{array}
[c]{cc}%
l_{\alpha\alpha} & 0\\
l_{\beta\alpha} & l_{\beta\beta}%
\end{array}
\right)  .
\]
Then\newline (a) $A_{DD|S_{q_{2}}}=L_{DD}\left(  L_{DD}\right)  ^{\prime}%
$\newline (b)$\left|  A_{DD|S_{q_{2}}}\right|  =\left(  l_{\alpha\alpha
}\right)  ^{2}\left(  l_{\beta\beta}\right)  ^{2}$\newline (c) $A_{\alpha
\alpha|S_{q_{2}}}=\left(  l_{\alpha\alpha}\right)  ^{2}$\newline (d)
$A_{\beta\beta|S_{q_{2}}}=\left(  l_{\beta\alpha}\right)  ^{2}+\left(
l_{\beta\beta}\right)  ^{2}$
\end{lemma}

\begin{proof}
The proof is straightforward and is omitted.
\end{proof}

Equation (\ref{like_ratio:g}) and parts (b)---(d)
of Lemma \ref{lemma_chol} give an efficient expression for the conditional
distributions in 
\sect\ref{sec:sampling}. \ The main computational effort is in updating
the Cholesky decompositions of the matrices $\Phi_{C_{q}C_{q}}$ and
$\Phi_{C_{q}C_{q}}^{\ast}$ whenever an edge is added or deleted. From Lemma
\ref{lemma_chol}, these Cholesky decomposition must be done with the entries
for the $i$th and $j$th vertices in the lower right corner. \ Note that
efficient Cholesky updating routines using Givens rotations are available in
Matlab and Fortran. \ Note also that the dimensions of $\Phi_{C_{q}C_{q}}$ and
$\Phi_{C_{q}C_{q}}^{\ast}$ depend on the cliques sizes and may be much
smaller than $p$. \ Thus our method has the local computational properties
described in \cite{Giudici_G99} and will have similar computational cost to 
their method per iteration of the Gibbs sampler.

%
\section*{Acknowledgement}
The research of Robert Kohn and Helen Armstrong was partially 
supported by an Australian Research Council Grant.

\end{document}